\documentclass[11pt]{article}
\pdfoutput=1
\usepackage{jcappub,natbib,float,caption,comment,bm,amsmath,wasysym}
\usepackage{ulem}
\usepackage[usenames,dvipsnames]{xcolor}

\usepackage{hyperref}

\definecolor{orange}{rgb}{1.0, 0.62, 0.0}

\def\be{\begin{equation}}
\def\ee{\end{equation}}
\def\ba#1\ea{\begin{align}#1\end{align}}
\newcommand{\vs}{\nonumber\\}

\newcommand{\refeq}[1]{equation~(\ref{eq:#1})}
\newcommand{\refeqs}[2]{equations~(\ref{eq:#1})--(\ref{eq:#2})}

\newcommand{\reffig}[1]{figure~\ref{fig:#1}}
\newcommand{\reffigs}[2]{figures~(\ref{fig:#1})--(\ref{fig:#2})}
\newcommand{\refFigs}[2]{Figures~(\ref{fig:#1})--(\ref{fig:#2})}
\newcommand{\refFig}[1]{Figure~\ref{fig:#1}}
\newcommand{\refsec}[1]{section~\ref{sec:#1}}

\newcommand{\refapp}[1]{appendix~\ref{app:#1}}

\newcommand{\vx}{\mbox{\boldmath$x$}}
\newcommand{\vr}{\mbox{\boldmath$r$}}
\newcommand{\vk}{\mbox{\boldmath$k$}}

\newcommand{\vv}{\mbox{\boldmath$v$}}
\newcommand{\vecs}{\mbox{\boldmath$s$}}

\newcommand{\hMpc}{~h^{-1}~{\rm Mpc}}
\newcommand{\ihMpc}{~h~{\rm Mpc}^{-1}}
\newcommand{\para}{\parallel}

\newcommand{\dd}{{\rm d}}
\newcommand{\cH}{{\cal H}}

\title{Generating Log-normal Mock Catalog of Galaxies in Redshift Space}
\author[a]{Aniket Agrawal,}
\author[a,b]{Ryu Makiya,}
\author[c]{Chi-Ting Chiang,}
\author[d,e]{Donghui Jeong,}
\author[a]{Shun Saito,}
\author[a,b]{Eiichiro Komatsu}
\affiliation[a]{Max-Planck-Institut f\"ur Astrophysik, Karl-Schwarzschild-Str. 1, 85741 Garching, Germany}
\affiliation[b]{Kavli Institute for the Physics and Mathematics of the
Universe, Todai Institutes for Advanced Study, the University of Tokyo,
Kashiwa, Japan 277-8583 (Kavli IPMU, WPI)}
\affiliation[c]{C.N. Yang Institute for Theoretical Physics, Department of Physics \& Astronomy,
Stony Brook University, Stony Brook, NY 11794, USA}
\affiliation[d]{Department of Astronomy and Astrophysics, The Pennsylvania State University, University Park, PA 16802, USA}
\affiliation[e]{Institute for Gravitation and the Cosmos, The Pennsylvania State University, University Park, PA 16802, USA}
\emailAdd{aniket@mpa-garching.mpg.de}
\abstract{
 We present a public code to generate a mock galaxy catalog in redshift
 space assuming a log-normal probability density function (PDF) of galaxy
 and matter density fields. We draw galaxies by Poisson-sampling the
 log-normal field, and calculate the velocity field from the linearised continuity
 equation of matter fields, assuming zero vorticity. This procedure
 yields a PDF of the pairwise velocity fields that is qualitatively
 similar to that of N-body simulations. We check fidelity
 of the catalog, showing that the measured two-point correlation
 function and power spectrum in real space agree with the input
 precisely. We find that a linear bias relation in the power spectrum
 does not guarantee a linear bias relation in the density contrasts,
 leading to a cross-correlation coefficient of matter and galaxies
 deviating from unity on small scales. We also find that linearising
 the Jacobian of the real-to-redshift space mapping provides a poor model
 for the two-point statistics in redshift space. That is, non-linear
 redshift-space distortion is dominated by non-linearity in the
 Jacobian. The power spectrum in redshift space shows a damping
 on small scales that is qualitatively similar to that of the well-known
 Fingers-of-God (FoG) effect due to random velocities, except that the
 log-normal mock does not include random velocities. This damping is a
 consequence of non-linearity in the Jacobian, and thus attributing the
 damping of the power spectrum solely to FoG, as commonly done in the
 literature, is misleading.
}
\begin{document}

\subheader{\rm YITP-SB-17-21}

\maketitle
\flushbottom

%%%%%%%%%%%%%%%%%%%%%%%%%%%%%%%%%%%%%%%%%%%%%%%%%%%%%%%%%%%%%%%%%%%%%%%%%%%%%
\section{Introduction}
\label{sec:intro}
%%%%%%%%%%%%%%%%%%%%%%%%%%%%%%%%%%%%%%%%%%%%%%%%%%%%%%%%%%%%%%%%%%%%%%%%%%%%%
Galaxy redshift surveys, mapping the three-dimensional distribution of
galaxies, have been one of the most powerful tools in modern cosmology
(see, for example, \cite{Percival2007}). Specifically, measurements of
the galaxy two-point correlation function or its Fourier counterpart,
power spectrum, allow us to extract cosmological information via, e.g.,
baryon acoustic oscillations %(BAO; \cite{Cole:2005sx,Eisenstein:2005su})
and the redshift-space distortion %(RSD; \cite{Turner:1977}). 
(BAO and RSD, see e.g.~\cite{Alam:2016hwk} for recent measurements).
Galaxy surveys complement other cosmological probes such as temperature
anisotropies and polarization of the Cosmic Microwave Background (CMB;
\cite{Bennett:2012zja,Adam:2015rua}) and luminosity distances of Type Ia
supernovae \cite{Conley:2011ku,Suzuki:2011hu}.

Deducing robust cosmological constraints from galaxy surveys requires an 
accurate modelling of the observed two-point correlation function and power 
spectrum along with their covariance matrices. This is a challenging 
task because of non-linearity and non-Gaussianity of the galaxy density field.
First, non-linear gravitational evolution transforms a nearly Gaussian 
initial density field into a non-Gaussian one \cite{Bernardeau:2001qr}, and the 
galaxy density field is related non-linearly to this non-Gaussian matter 
density field ({\it galaxy bias}; see \cite{Desjacques:2016bnm} for a
review). In addition, the observed galaxy density field differs from the
underlying one because of systematics due to peculiar velocity (RSD) and
variations in observing conditions across the survey area 
(window function effect).

Due to these various non-linearities, unlike for the
CMB analysis, the Gaussian approximation is no longer valid for
computing the covariance matrix of the galaxy two-point
statistics. Going beyond the Gaussian approximation, a method based on
non-linear perturbation theory including contributions from connected
four-point functions can model the non-linear covariance matrix on
quasi-linear scales
\cite{Mohammed:2016sre,Barreira:2017sqa}. Perturbative approaches break
down on small scales where non-linearities are too strong. The
gravitational amplification and galaxy bias in these non-linear scales
may be fitted by a number of free parameters of effective field theory
\cite{baumann/etal:2012,carrasco/etal:2012}. However, treatment of
the mode-coupling effect due to the survey window function (for example, due
to sparse sampling of the survey area \cite{chiang/etal:2013}) requires
a full account of modes down to the resolution scale of the survey, set
by the number density of the sample: $\bar{n}_g\gtrsim1/P(k)$, $P(k)
\equiv \left| \delta(\vk)\right|^2$ being the power spectrum of the
overdensity field $\delta(\vx)$.

Cosmological N-body simulations have been the gold standard in modelling 
non-linearity in the large-scale structure. As phenomena in a wide range of 
scales are involved in the formation and evolution of galaxies, simulating all 
the relevant physics of the formation and evolution of galaxies is impractical.
Instead, the usual practice is to ``paint'' galaxies onto the halos in 
matter-only simulations by using the halo-occupation distribution (HOD) 
function estimated from, for example, the angular clustering of the survey 
(e.g., \cite{Guo:2015dda}), or by using Subhalo Abundance Matching 
(SHAM) with the observed stellar mass function (e.g., \cite{Saito:2015eka}.
With these {\it mock} galaxy samples from the simulation 
at hand, the galaxy correlation functions and their covariance matrices can be 
measured directly from a suite of N-body simulations including various 
selection effects of the surveys. Even for these matter-only N-body
simulations, however, a robust cosmological parameter estimation may
demand too large computational resources. This is because estimating the
covariance matrix from N-body simulations hampers the cosmological
parameter estimation by a factor of $1+N_b/N_s$, where $N_s$ is the
number of N-body simulations and $N_b$ is the number of independent
bins used for the estimation of parameters \cite{Dodelson:2013uaa}. 
If we were to achieve a percent precision on the covariance matrix, we
would need $N_b/N_s=10^{-2}$. 
As $N_b\approx 10^2$ for typical survey data, $N_s\approx 10^4$ would be required.
This requirement would become more severe in estimating the inverse covariance matrix 
and its associated errors (see e.g., \cite{Hartlap:2006kj,Percival2007,Sellentin:2016psv}).

One pragmatic way of bypassing this problem is to simulate gravitational 
evolution by adopting a set of simplified assumptions. In this approach,
one trades accuracy for speed of simulations, especially on small 
scales. For example, the Zel'dovich simulation \cite{zeldovich:1970} captures 
correct density and velocity fields on large scales where
non-linearities are modest; the higher order Lagrangian perturbation theory 
(LPT \cite{moutarde/etal:1991,buchert/ehlers:1993,catelan:1995}) 
simulations capture non-linearities on progressively smaller scales 
\cite{coles/melott/shandarin:1993,melott/buchert/weib:1995,kitaura/hess:2013,tassev/etal:2013,neyrinck:2016,koda/etal:2016}.
We refer the readers to Ref.~\cite{monaco:2016} for a recent review and to 
Refs.~\cite{melott:1994,Chuang:2014toa,munari/etal:2017} for comparisons
between different approaches.

In this paper, we shall take a different approach: instead of modelling
the non-linear density evolution, we exploit statistical properties of
the non-linear galaxy density field. Specifically, we generate a mock
galaxy catalog with the assumption 
that the probability density function (PDF) of galaxy density fields
follows a log-normal distribution. This assumption is based upon the
observation that the PDF of log-transformed density fields, 
$\ln(1+\delta)$ with $\delta\equiv n/\bar{n}-1$ being the density contrast, 
measured from N-body simulations roughly matches a Gaussian PDF
\cite{coles:1991,colombi:1994,kofman:1994,bernardeau/kofman:1995,
uhlemann/etal:2016,shin/etal:2017}. The evidence for a log-normal PDF does
not only come from the matter density fields in simulations, but also
from the Dark Energy Survey (DES) science verification data
\cite{Clerkin:2016kyr} and earlier measurements \cite{Hubble:1934,Wild:2004me}. 

Note that log-normality is not merely a
statement about the one-point PDF, but it means that the log-transformed
field $\ln(1+\delta)$ is a multi-variate Gaussian random field whose
statistics are completely specified by its two-point correlation
function. For this, the N-body simulation of Ref.~\cite{Kayo:2001gu}
has confirmed that the two-point correlation function of matter density
fields also roughly matches the prediction of log-normality well into
fairly non-linear regime.

In addition, the log-normal mock generator presents the following practical 
advantages that further motivate our pursuing this approach: 
\begin{enumerate}
{\item
It is fast.
Since the relation between the density fields and the Gaussian 
(log-transformed) fields is given by a local transformation 
(see \refsec{simulation} for more details), the log-normal mock generator is
almost as fast as generating three-dimensional Gaussian random
      fields. This allows us to quickly generate a large number of mock galaxy 
distributions.}
{\item
It is direct. The log-normal mock generator takes the observed galaxy 
two-point correlation function as an input so that we can avoid 
post-processing steps (halo finding, HOD, for example) connecting
the non-linear density field to mock galaxies.}
{\item
It is instructive. Upon assuming log-normal PDF of the galaxy density
      field, all higher-order correlation functions are given in terms
      of the two-point correlation function of the log-transformed field
      \cite{coles:1991}. This allows us to quantitatively study highly 
      non-linear mode-coupling effects in both the signal and covariance matrix
      that demand knowledge about the density
      field on non-linear scales. One such example is mode-coupling due
      to the survey window function. By using a thousand log-normal mock
      catalogs, Ref.~\cite{chiang/etal:2013} has quantified the effect
      from a duplicated, sparse (instead of contiguous) angular
      selection function, and deduced the optimal analysis strategy.
}
\end{enumerate}

In this paper, we extend the real-space log-normal mock generator
presented in Ref.~\cite{chiang/etal:2013} by including the velocity field in a
consistent manner. We then generate the log-normal mock in
redshift space by applying the real-to-redshift space mapping. Again, equipped
with perfect knowledge about the statistical properties of the galaxy
density and velocity fields, such a mock catalog serves as an excellent
test bed for modelling RSD due to this non-linear mapping \cite{scoccimarro:2004}.
To test the RSD effect on the two-point statistics of the log-normal 
mock catalog, we begin with the real-space galaxy two-point correlation 
function as an input. We use a log-normal PDF to generate a
three-dimensional galaxy density field, as well as a matter density
field. Finally, we generate a velocity field consistent with the matter
density field by using the linearised continuity equation (see
\refsec{simulation} for more details). We then measure the galaxy
two-point statistics (correlation function and power spectrum) both in
real and redshift space, and the pairwise line-of-sight velocity PDFs
from the log-normal mock catalog. We also calculate the mean pairwise
velocity using log-normal statistics and show that it agrees with the
measurement from the catalog.

Our implementation of log-normal galaxy density and velocity fields
differs from other log-normal codes such as \cite{Pearson:2016jzc}, FLASK \cite{Xavier:2016elr}
and CoLoRe \cite{Alonso:2014sna}. FLASK does not have a prescription for
producing a velocity field; hence one has to provide an anisotropic
power spectrum when generating density fields in redshift space. CoLoRe
generates a velocity field by using linear theory velocities
corresponding to the log-transformed field, so the resulting velocities
follow a Gaussian PDF. Note that the fact that the velocity field
follows a Gaussian PDF does not imply that the pairwise line-of-sight
velocity PDF is Gaussian because the pairwise line-of-sight velocity PDF
is a pair-weighted quantity (see \refsec{review} for more details). In
contrast, in this paper, we use the linearised continuity equation to
ensure mass conservation with little additional computing cost compared
to the CoLoRe method.

The rest of the paper is organized as follows. In \refsec{review} we
present an overview of redshift-space statistics including the two-point
correlation function and the pairwise line-of-sight velocity PDF. In
\refsec{simulation} we introduce our method to generate log-normal
density and velocity fields. In \refsec{code_performance} we present
measurements from our log-normal mock catalogs, including two-point
statistics in real space (sec.~\ref{sec:real_space_2pt}), the
cross-correlation coefficient between matter and galaxy fields
(sec.~\ref{sec:real_space_cxi}), two-point statistics in redshift space
(sec.~\ref{sec:zspace_stat}), and pairwise line-of-sight velocity PDFs
(sec.~\ref{sec:real_space_pdf}); in sec.~\ref{sec:kaiser_limit} we
discuss how the streaming model reduces to the Kaiser limit at large separations.
We summarize the results in \refsec{summary}.
In \refapp{streaming}, we present a derivation of the streaming model
for RSD. In \refapp{pk_bin}, we lay out the method that we use to
correct for the binning effect when measuring power spectra. In
\refapp{mean_pairwise_v}, we present the details of calculating the mean
pairwise line-of-sight velocity from the log-normal mock catalogs. The
code documentation of our log-normal mock generator\footnote{The mock
generation code is publicly available as ``lognormal\_galaxies'' at
\url{http://wwwmpa.mpa-garching.mpg.de/\textasciitilde
komatsu/codes.html}.} is in \refapp{docu}. Throughout, we use the
following Fourier convention:
\be
f(\vk) = \int d^3x f(\vx) e^{-i\vk\cdot\vx},~
f(\vx) = \int \frac{d^3k}{(2\pi)^3} f(\vk) e^{i\vk\cdot\vx}\,.
\ee
In this paper, the term {\it real} space refers to the contrast with redshift 
space, and the term {\it configuration} space refers to the contrast with 
Fourier space.

%%%%%%%%%%%%%%%%%%%%%%%%%%%%%%%%%%%%%%%%%%%%%%%%%%%%%%%%%%%%%%%%%%%%%%%%%%%%%
\section{Review of RSD}
\label{sec:review}
%%%%%%%%%%%%%%%%%%%%%%%%%%%%%%%%%%%%%%%%%%%%%%%%%%%%%%%%%%%%%%%%%%%%%%%%%%%%%
In spectroscopic galaxy redshift surveys, radial distances to galaxies
are inferred from observed spectral shifts containing both the Hubble
expansion and peculiar velocities of galaxies along the
line-of-sight. The observed positions (redshift space) of galaxies are
related to the true positions (real space) of galaxies by
\be
 \vecs=\vx+\frac{1}{\mathcal{H}}\vv(\vx)\cdot \hat{\ell} \,.
\label{eq:r_to_s}
\ee
Here, $\vx$ and $\vecs$ are the comoving coordinates, respectively, in real 
and redshift space, $\mathcal{H}$ is defined by $\mathcal{H}\equiv aH$ with $H$ being the Hubble expansion rate 
and $a$ being the scale factor of the universe, $\vv=d\vx/d\eta$ is the peculiar 
velocity of the galaxy with $\eta$ being the conformal time (related to the time coordinate by $d\eta=dt/a(t)$), and 
$\hat{\ell}$ is the line-of-sight direction of the galaxy. In this paper, we 
shall take the plane-parallel (distant-observer) approximation such that 
$\hat{\ell}\equiv\hat{z}$ is fixed for all galaxies in the survey.

As a result of the shift in the line-of-sight distance given by
\refeq{r_to_s}, the observed galaxy distribution in redshift space is
anisotropically distorted from the underlying real-space
distribution. It is anisotropic because the distortion happens only
along the line-of-sight direction. Since the number of galaxies in real
and redshift space must be the same,
we have
\be
 [1+\delta_g^s(\vecs)]d^3s=[1+\delta_g(\vx)]d^3x \,,
\label{eq:conservation}
\ee
where $\delta_g\equiv n_g/\bar{n}_g-1$ and 
$\delta_g^s\equiv n_g^s/\bar{n}_g^s-1$ are the galaxy density contrasts in real 
and redshift space, respectively. Here, we ignore the time evolution of 
the mean number density of galaxies so that $\bar{n}_g=\bar{n}_g^s$;
this would induce the evolution bias ($b_e$) contribution in 
Ref.~\cite{jeong/schmidt/hirata:2012} which is small for $kr\gg1$.
We then relate the redshift-space density contrast $\delta^s_g$ to the real 
space one $\delta_g$ as 
\be
\delta_g^s(\vecs) = [1+\delta_g(\vx)]J(\vx) - 1,
\label{eq:dgs_dgJ}
\ee
with the Jacobian of the coordinate transformation
\be
 J(\vx)
=
\left|\frac{d^3x}{d^3s}\right|
=\left[1+\frac{1}{\mathcal{H}}\frac{\partial v_z(\vx)}{\partial z}\right]^{-1} \,,
\label{eq:jacobian}
\ee
where $z$ refers to the line-of-sight coordinate. Note that the
relation above only works 
when the distant-observer approximation is valid and
when the real-to-redshift coordinate mapping 
[\refeq{r_to_s}] is one-to-one; otherwise, the Jacobian would be
infinite. 

There are two sources of non-linearity in the relationship between real-
and redshift-space density contrasts. First, the Jacobian of the
mapping, \refeq{jacobian}, is a non-linear function of the velocity
field. Second, the velocity field itself is non-linear due to
gravitational evolution at late times. Ignoring the velocity bias
\cite{Desjacques:2016bnm} that only affects at galaxy formation scales,
we assume that the peculiar velocity is sourced by the underlying matter
density fluctuation and that galaxies are moving with the same velocity
as matter. With this assumption, the peculiar velocity field is governed
by the continuity equation for matter density contrast
\be
\frac{\partial\delta_m(\vx)}{\partial\eta}+\nabla\cdot\lbrace\left[1+\delta_m(\vx)\right]\vv(\vx)\rbrace=0 \,,
\label{eq:continuity}
\ee
and the Euler equation, 
\be
 \frac{\partial\vv(\vx)}{\partial\eta}+\mathcal{H}\vv(\vx)+\vv(\vx) \cdot \nabla \vv(\vx) = -\frac{3}{2}\mathcal{H}^2\nabla^{-1}\delta(\vx) \,.
\ee

In the large-scale limit, in which the density contrast and the peculiar 
velocity are small, we can linearise both the Jacobian and the continuity 
equation to obtain
\be \label{eq:cont_eq_k}
 \delta_g^s(\vecs)=b \delta^{L}_{m}(\vx)-\frac{1}{\mathcal{H}}\frac{\partial v_z(\vx)}{\partial z} \,, \quad
 \vv(\vk)=i\mathcal{H}f\frac{\vk}{k^2}\delta^{L}_{m}(\vk) \,,
\ee
where $\delta^{L}_{m}$ is the linear matter density contrast, 
$f=\dd \ln D/\dd\ln a$ is
the logarithmic growth rate with $D$ being the linear growth factor, and $b$ 
is the linear bias factor. Then the linear redshift-space galaxy 
power spectrum $P_{gg}^s(k,\mu_k)$ becomes 
\be
 P_{gg}^s(k,\mu_k)=\left(b+f\mu_k^2\right)^2 P^{L}_{m}(k) \,,
\label{eq:kaiser_pk}
\ee
where $\mu_k=\hat{k}\cdot\hat{z}$
is the cosine of the angle between the line-of-sight and the wave vector $\vk$,
and $P^{L}_{m}$ is the linear matter power spectrum. This is the
so-called Kaiser formula \cite{kaiser:1987}. It is useful to expand
the redshift-space power spectrum using the Legendre polynomials
$\mathcal{L}_\ell(\mu_k)$,
\be
P_{gg}^s(k,\mu_k) = \sum_{\ell} P_{gg,\ell}^s(k)\mathcal{L}_\ell(\mu_k).
\label{eq:multipole}
\ee
Non-zero components are monopole, quadrupole, and hexadecapole, which
are given respectively by
\ba
 P_{gg,\ell=0}^s(k)
\:=&\left(b^2+\frac23bf+\frac15f^2\right)P^{L}_{m}(k) \,, \vs
 P_{gg,\ell=2}^s(k)
\:=&\left(\frac43bf+\frac47f^2\right)P^{L}_{m}(k) \,, \vs
 P_{gg,\ell=4}^s(k)
\:=&\frac{8}{35}f^2P^{L}_{m}(k) \,.
\ea
The corresponding galaxy two-point correlation function is given in a similar
manner as \cite{hamilton:1992,hamilton:1998}:
\be
\xi_{gg}^s(s,\mu) 
=
\left(b^2 + \frac23 bf + \frac15 f^2\right) \xi_0(s)
-
\left(\frac43 bf + \frac47 f^2\right) {\cal L}_2(\mu) \xi_2(s)
+
\frac{8}{35}f^2 {\cal L}_4(\mu) \xi_4(s),
\ee
with 
\be
\xi_\ell(s) \equiv \int \frac{dk}{2\pi^2} k^2 P_m^L(k) j_\ell(ks)\,.
\label{eq:def_xils}
\ee

On very small scales, corresponding to the interior of virialized
objects such as galaxy clusters, peculiar velocities are randomly
oriented. As a result, the clustering amplitude is reduced along the
line-of-sight; this effect is called Fingers-of-God (FoG;
\cite{Jackson:2008yv}), as clusters appear elongated along the
line-of-sight direction. The small-scale damping of the power spectrum
due to FoG is often modelled by introducing an exponential or a
Lorentzian damping factor motivated by the pairwise line-of-sight
velocity PDF measured from N-body simulations \cite{scoccimarro:2004}.

On intermediate scales, the Jacobian and the continuity equation cannot 
be linearised, and galaxies are not in random motion in virialised objects. We 
thus need to take into account the non-linear effects in the velocity field as 
well as in the Jacobian. Modelling non-linear RSD has been studied
extensively in the literature for the past decade including, for
example, standard (Eulerian) perturbation theory
\cite{Heavens:1998es,Bernardeau:2001qr}, Lagrangian perturbation theory
\cite{Matsubara:2007wj,Matsubara:2008wx,Matsubara:2011ck,Matsubara:2013ofa,Sugiyama:2013mpa},
effective field theory \cite{Lewandowski:2015ziq,Perko:2016puo}, and the
distribution function approach
\cite{Seljak:2011tx,Okumura:2011pb,Okumura:2012xh,Vlah:2012ni,Vlah:2013lia,Okumura:2015fga}. All
these methods are based on non-linear perturbation theory and treat both
non-linearities in the velocity field and the Jacobian
perturbatively. The resummation approaches
\cite{Taruya:2010mx,Wang:2013hwa}, and the streaming model
\cite{peebles:book,Davis:1982gc,Fisher:1994ks,scoccimarro:2004,uhlemann:2015}, 
on the other hand,
can accommodate the full non-linearities in the Jacobian. Here, we focus
on the streaming model.

The streaming model describes RSD in the galaxy 
two-point correlation function as a mapping between galaxy pairs in real
and redshift space. This method aligns well with the interpretation that the 
galaxy two-point correlation function is the excess number of pairs over the 
cosmic mean. Mathematically, denoting the 
pairwise line-of-sight velocity PDF as $\mathcal{P}(s_\para-r_\para,\vr)$, 
that is, in terms of the change in the line-of-sight separation
$r_\para-s_\para \equiv - \Delta v_{z}/\mathcal{H}$, 
the redshift-space galaxy two-point correlation function $\xi_{gg}^s$ 
can be written as 
\be
 1+\xi_{gg}^s(s_\para,s_\perp)=\int dr_\para~\left[ 1+\xi_{gg}(r)\right] 
 \mathcal{P}(s_\para-r_\para,\vr) \,,
\label{eq:streaming}
\ee
where $\xi_{gg}$ is the real-space galaxy correlation function. 
We show a derivation of the streaming model in \refapp{streaming},
assuming only number conservation and statistical
homogeneity of the Universe. Once the pairwise line-of-sight
velocity PDF $\mathcal{P}(s_\para-r_\para, \vr)$ is known accurately,
one can map the real-space correlation function into redshift space by
\refeq{streaming}. Of course, the linearised streaming model reproduces
the linear theory result \cite{Fisher:1994ks,jeong/etal:2015}.

\begin{figure}[t]
\centering
\includegraphics[width=1\textwidth, bb=0 0 500 327]{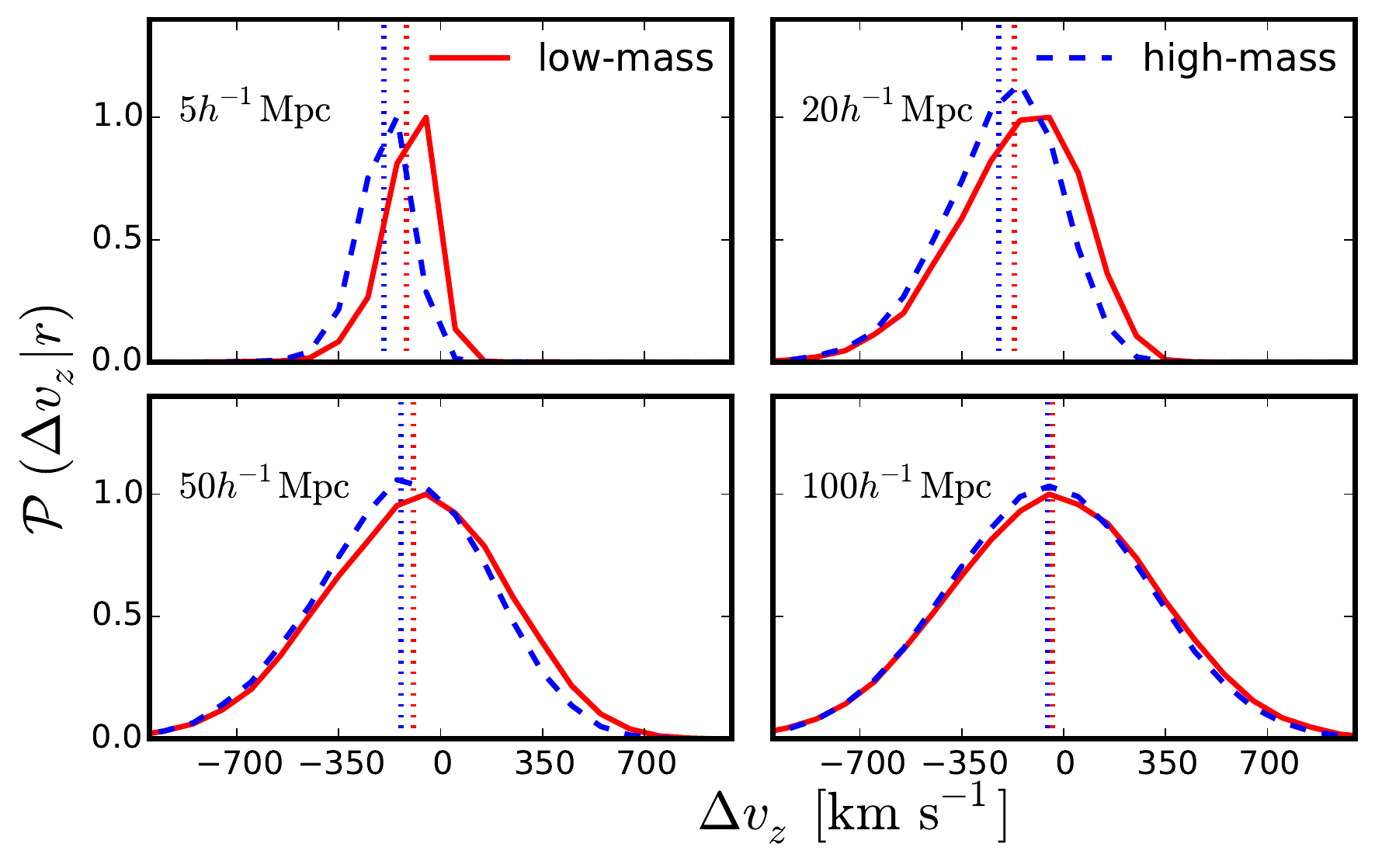}
\caption{
Pairwise line-of-sight velocity PDFs averaged over 160 N-body
 simulations along the line-of-sight ($0.99\leq\mu\leq1.0$) for four
 separations of galaxy pairs: $5.25\hMpc$ (top left), $20.25\hMpc$ (top
 right), $50.25\hMpc$ (bottom left), and $100.25\hMpc$ (bottom
 right). The simulation volume is $(2400\hMpc)^3$ and the output
 is at $z=0$. The red solid and blue dashed lines denote the PDFs for
 low-mass ($5 \times 10^{13} - 6.5 \times 10^{13}$ $h^{-1}\,M_{\odot}$)
 and high-mass ($1 \times 10^{14} - 2.5 \times 10^{14}$
 $h^{-1}\,M_{\odot}$) halos, for which the average halo biases are
 $b=1.8$ and 2.5, respectively. Negative (positive)
 velocities denote galaxy pairs moving towards (away from) each
 other. The red and blue dashed vertical lines denote the mean pairwise
 line-of-sight velocity for low-mass and high-mass halos
 respectively. The PDFs have been normalised to unity, and rescaled such
 that the maximum of the red lines is unity. For small separations (top
 panels), there is a small but significant shift towards more negative
 values for PDFs with a higher mass; for larger separations (bottom
 panels), the PDFs become more symmetric. However, the mean velocity is
 still not zero even at separations of $\sim 100 h^{-1}\,\rm Mpc$.}
\label{fig:pdf_nbody}
\end{figure}

The key characteristic of the pairwise line-of-sight velocity PDF is
that it is a \textit{pair-weighted}
quantity. Ref.~\cite{scoccimarro:2004} (also see \refapp{streaming})
shows the moment generating function of the pairwise
line-of-sight velocity PDF as
\begin{eqnarray}\label{eq:mgf_pdf}
 \mathcal{P}(s_\parallel-r_\parallel, \vr) & = &  
\int \frac{d\gamma}{2\pi}e^{i\gamma (s_\parallel-r_\parallel)}
 \mathcal{M}(-i\gamma, \vr),\\
 \mathcal{M}(\lambda, \vr) & = & 
 \frac{\left\langle e^{\lambda (v_{z}(\vx_1)-v_{z}(\vx_2))/\cH}
 \left[ 1+\delta_{g}(\vx_1)\right] \left[ 1+\delta_{g}(\vx_2)\right] \right\rangle}{1+\xi_{gg}(r)},
\end{eqnarray}
where $\vr = \vx_1-\vx_2$.
We show in \reffig{pdf_nbody} the pairwise line-of-sight velocity PDF of dark 
matter halos averaged over 160 N-body simulations
\cite{deputter:2011}. The box size is $2400\hMpc$ on a side, and
the redshift is $z=0$.
We find that the PDF has a negative mean (vertical dotted lines) and a 
negative skewness, and the trend is more obvious for smaller separations. 
In our sign convention, this means that there are more approaching pairs than 
recessing pairs, which is a consequence of the attractive nature of gravity.
For larger separations, linear theory applies so that both the mean 
pairwise velocity and the skewness get smaller, and the distribution becomes 
more symmetric. We also find that the mean of the PDF is more negative for 
larger mass halos. 

%%%%%%%%%%%%%%%%%%%%%%%%%%%%%%%%%%%%%%%%%%%%%%%%%%%%%%%%%%%%%%%%%%%%%%%%%%%%%
%%%%%%%%%%%%%%%%%%%%%%%%%%%%%%%%%%%%%%%%%%%%%%%%%%%%%%%%%%%%%%%%%%%%%%%%%%%%%
Several attempts have been made in the literature 
\cite{bianchi:2016,uhlemann:2015,wang/reid/white:2013,reid:2011}
to model the redshift-space galaxy two-point correlation 
function by calculating ${\cal P}(s_\para-r_\para,\vr)$ analytically.
It is still difficult to predict redshift-space two-point correlation 
function on all scales without the aid of free parameters. 
The key issue in modelling the pairwise line-of-sight velocity PDF 
is its non-Gaussianity; as shown in Ref.~\cite{scoccimarro:2004,Fisher:1994ks},
the PDF is non-Gaussian even when the density and velocity field follow a
Gaussian distribution. As discussed in \refsec{real_space_pdf}, we also find this
non-Gaussianity in the log-normal mock catalogs. The pairwise
line-of-sight velocity PDFs that we measure from the log-normal catalogs
show qualitatively the same features as those from N-body simulations. In both cases we recover the Kaiser limit on large scales. In \refsec{kaiser_limit} we quantitatively discuss how the streaming model reduces to the Kaiser limit at large separations, for our log-normal mocks, making use of the moments of the measured pairwise line-of-sight velocity PDF. 

%%%%%%%%%%%%%%%%%%%%%%%%%%%%%%%%%%%%%%%%%%%%%%%%%%%

\section{Log-normal Catalog Generation}
\label{sec:simulation}
The log-normal distributed density contrast $\delta(\vx)$ is related to a 
Gaussian (log-transformed) field 
$G(\vx)\equiv \ln\left[1+\delta(\vx)\right] - \left<
\ln\left[1+\delta(\vx)\right]
\right>$ as 
\be
 \delta(\vx)=e^{-\sigma_G^2+G(\vx)}-1 \,,
\label{eq:delta_G}
\ee
where the pre-factor with the variance of the Gaussian field 
$\sigma_G^2\equiv\langle G^2\rangle$ ensures that the mean of $\delta(\vx)$ 
vanishes. Note that the log-normal density fields follow the natural 
constraint $\delta(\vx)\geq-1$ of density contrasts by definition. 
This is not the case for simulations that generate the linear density 
contrast from Gaussian realisations, although the violation rarely occurs 
when the variance is small at, e.g., high redshift for setting up 
the initial conditions of simulations.
Applying \refeq{delta_G}, one can relate the two-point correlation function of 
the Gaussian field $\xi^G(r)$ to the two-point correlation function of the 
density field $\xi(r)$ as 
\cite{coles:1991}
\be
 \xi^{G}(r) = \text{ln}\left[ 1+\xi(r)\right] \,.
\label{eq:xiG and xi}
\ee

Since Gaussian fields of different Fourier modes are uncorrelated, we 
generate $G$ in Fourier space. To generate a 
log-normal density field with a given power spectrum $P(k)$, we first Fourier 
transform the power spectrum to get the target two-point correlation function 
$\xi(r)$. We then calculate the two-point correlation function of the Gaussian 
field $\xi^G(r)$ by \refeq{xiG and xi}, and Fourier transform $\xi^G(r)$ to 
get the power spectrum $P^G(k)$ of $G$.  
The Fourier space Gaussian field $G(\vk)$ is generated with 
\cite{djeong_thesis:2010}
\begin{equation}
 G(\vk)=\sqrt{\frac{P^G(k)V}{2}}\left(\theta_r + i\theta_i \right)\,,
\label{eq:Gk}
\end{equation}
where $\theta_r$ and $\theta_i$ are Gaussian random variables with unit 
variance and zero mean, and $V$ is the volume of the simulation. 
We also enforce $G(-\vk)=G^*(\vk)$ so that the 
Gaussian field in configuration space $G(\vx)$ is real.
After $G(\vk)$ is generated at each point in the Fourier grid,
we use FFTW library \cite{fftw} to Fourier-transform $G(\vk)$ and obtain 
$G(\vx)$ on regular cells in configuration space. We then use \refeq{delta_G} 
to transform $G(\vx)$ into the desired log-normal density contrast 
$\delta(\vx)$ on each cell, with the variance $\sigma_G$ measured from 
$G(\vx)$ in all cells. 
The resulting density fluctuation $\delta(\vx)$ follows a log-normal 
distribution with the target power spectrum $P(k)$. 

At each cell in configuration space, we calculate the expectation value for the number of 
galaxies $N_g(\vx) = \bar{n}_g[1+\delta(\vx)]V_{\rm cell}$, where $\bar{n}_g$ 
is the global mean galaxy number density and $V_{\rm cell}$ is the volume of 
the cell. As $N_g(\vx)$ is not an integer, we draw a Poisson random number with the mean $N_g(\vx)$ to obtain the integer number of galaxies in the cell and populate galaxies
randomly within the cell. This discretisation is consistent with the
nearest-grid-point (NGP) density assignment in the sense that the
galaxies are equally spread over the cell.

We next assign velocities to galaxies. For our mock catalogs, we estimate 
velocities by using the linearised continuity equation of the matter fields:
\be
 \frac{\partial\delta_m(\vx)}{\partial\eta} + \nabla \cdot \vv(\vx) = 0 \,, \quad {\rm or} \quad
 \vv(\vk)=i\mathcal{H}f\frac{\vk}{k^2}\delta_m(\vk) \,.
\label{eq:cont_eq_x}
\ee
As the velocity bias can be ignored at the leading order
\cite{Desjacques:2016bnm}, the velocity of a galaxy follows the local
matter velocity. We implement this equation as follows.
We take the target matter power spectrum to compute $\delta_m(\vk)$
on Fourier cells following the procedures described above, use 
\refeq{cont_eq_x} to compute $\vv(\vk)$ on each Fourier grid, and then 
Fourier-transform $\vv(\vk)$ back into configuration space to obtain 
$\vv(\vx)$ on each cell. To ensure that the galaxy overdensities and 
velocities are correlated, we use the same random seed for $G$ of galaxies and matter. Namely, the phases of $G_g(\vk)$ and $G_m(\vk)$ are identical; however, this does not imply that the phases of $\delta_g(\vk)$ and $\delta_m(\vk)$ are identical, as we show in \refsec{real_space_cxi}.
Finally, we assign the same velocity to all galaxies within one cell.

The target galaxy and matter power spectra can be chosen freely. We need
a galaxy bias model \cite{Desjacques:2016bnm} to find the matter power
spectrum that is consistent with the chosen galaxy power spectrum. In this paper, we
use a linear bias relation between the matter and galaxy power spectra,
$P_{gg}(k) = b^2 P_{mm}(k)$, with the linear bias parameter $b$. One
important feature of the log-normal catalogs is that even though the
target galaxy and matter power spectra are linearly related, the density
fields are not proportional to each other (see
\refsec{real_space_cxi}). Also, while the galaxy power spectrum and the
matter power spectrum are linearly related, the power spectra of their
corresponding Gaussian fields are not proportional to each other because
\begin{equation}
 \ln[1+b^2\xi(r)] \neq b^2 \ln[1+\xi(r)] \,.
\end{equation}

We generate 50 log-normal mock catalogs in a cubic volume with 
$L_{\rm box} = 1000\hMpc$, and $1024^3$ grids for the Fourier transformation.
This corresponds to the Nyquist frequency of $k_{Ny} = 3.22\ihMpc$.
The catalogs are generated at $z=1.3$, and each catalog contains roughly 2.1
million galaxies. We compute the input galaxy power spectrum from the 
linear matter power spectrum using Eisenstein and Hu's fitting function 
\cite{eisenstein/etal:1997} and the linear galaxy bias $b=1.455$. We assume a flat $\Lambda$CDM model with $\Omega_m = 0.272$, $n_s = 0.963$, $A = 2.1 \times 10^{-9}$. 
The outcome of this mock generator is a set of positions and velocities of 
galaxies in three-dimensional space with the target galaxy power spectrum.
We shall present detailed tests on the output catalogs in 
\refsec{code_performance}.

%%%%%%%%%%%%%%%%%%%%%%%%%%%%%%%%%%%%%%%%%%%%%%%%%%%%%%%%%%%%%%%%%%%%%%%%
\section{Validation of the Log-normal Mocks}
\label{sec:code_performance}
%%%%%%%%%%%%%%%%%%%%%%%%%%%%%%%%%%%%%%%%%%%%%%%%%%%%%%%%%%%%%%%%%%%%%%%%
In this section, we present the results of the log-normal mock generator.
We start from the two-point statistics in real space (\refsec{real_space})
and then move onto the redshift space correlation function 
(\refsec{zspace_stat}), the pairwise velocity PDF 
(\refsec{real_space_pdf}) and recovery of the Kaiser limit on large scales (\refsec{kaiser_limit}).

%%%%%%%%%%%%%%%%%%%%%%%%%%%%%%%%%%%%%%%%%%%%%%%%%%%%%%%%%%%%%%%%%%%%%%%%
\subsection{Real-space density statistics}
\label{sec:real_space}
%%%%%%%%%%%%%%%%%%%%%%%%%%%%%%%%%%%%%%%%%%%%%%%%%%%%%%%%%%%%%%%%%%%%%%%%
\subsubsection{Two-point Statistics}
\label{sec:real_space_2pt}
%%%%%%%%%%%%%%%%%%%%%%%%%%%%%%%%%%%%%%%%%%%%%%%%%%%%%%%%%%%%%%%%%%%%%%%%
We first measure the real-space two-point statistics: power spectrum and
two-point correlation function. As we have pointed out earlier, our
log-normal catalogs populate galaxies randomly in each cell, and this is
equivalent to adopting the NGP mass assignment scheme. To be consistent,
we use NGP with the same grid number to estimate the galaxy density
contrast for Fast Fourier Transform (FFT). In this way, we recover the
input target power spectrum (having subtracted a constant shot noise $=1/\bar{N}$, 
$\bar{N}$ denoting the number density of galaxies), without needing to deconvolve 
the window function due to the density assignment \cite{jing:2004}.
Should we use a different mesh number or density assignment scheme 
(such as Cloud-In-Cell), we would have to correct for the window function 
effects by applying an appropriate deconvolution.

\begin{figure}[t]
\centering
\includegraphics[width=0.9\textwidth, bb = 0 0 517 332]{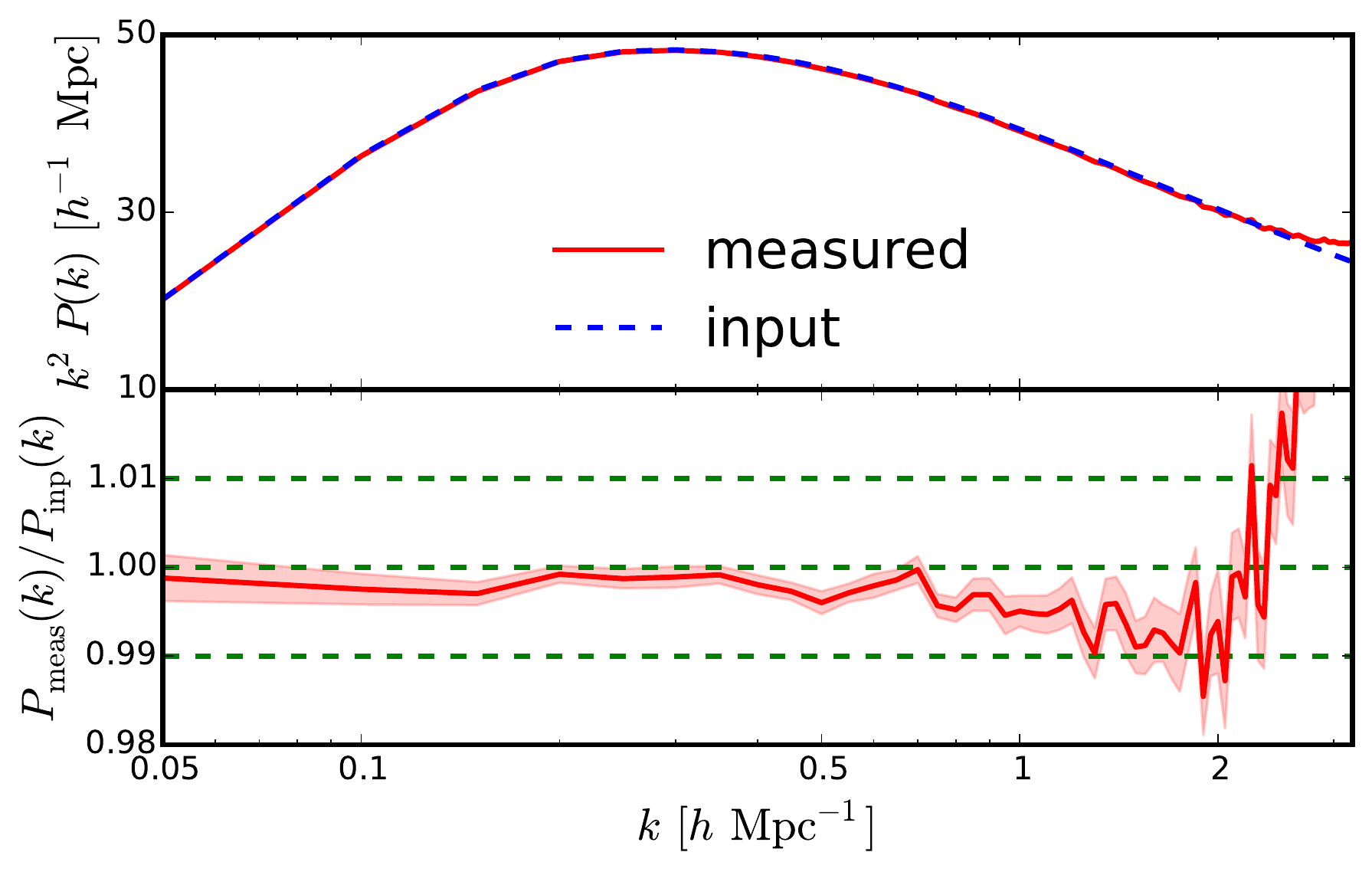}

\caption{
(\textit{Top}) Mean of the real-space galaxy power spectrum measured
 from 50 log-normal catalogs (solid) and the input power spectrum
 (dashed). We show $k^2P(k)$. (\textit{Bottom}) Ratio of the two. The band shows the error
 on the mean estimated from 50 realisations. The Nyquist frequency for these measurements is $k_{\rm Ny} = 3.22\, h$ Mpc$^{-1}$.
 }
\label{fig:pk_real}
\end{figure}

\refFig{pk_real} shows the comparison between the input power spectrum and the 
power spectrum averaged over 50 log-normal realisations (top), and the
ratio of the two (bottom). The band shows the error on the mean estimated from 50 realisations. We find an excellent agreement between the measured and
input power spectra for $k\lesssim2\ihMpc$. Note that, when comparing the
measurement and prediction we need to take special care to calculate the
correct {\it effective} wavenumber by which each bin of the measured power 
spectrum is represented; this is because the binned power spectrum is averaged 
over many different wavenumbers that fall into the binning criteria.
This effect is particularly important on large scales where the number of
Fourier modes is small. We present the details of this correction 
in \refapp{pk_bin}. %Alternatively, one can average over the wavevectors 
%included in each bin to find an effective wavenumber 
%\cite{sefusatti/crocce/desjacques:2010}, but it turns out that the method  
%of discretizing the large-scale modes is the most accurate.

%%%%%%%%%%%%%%%%%%%%%%%%%%%%%%%%%%%%%%%%%%%%%%%%%%%%%%%%%%%%%%%%%%%%%%%%
%%%%%%%%%%%%%%%%%%%%%%%%%%%%%%%%%%%%%%%%%%%%%%%%%%%%%%%%%%%%%%%%%%%%%%%%
\begin{figure}[t]
\centering
\includegraphics[width=0.8\textwidth, bb = 0 0 527 332]{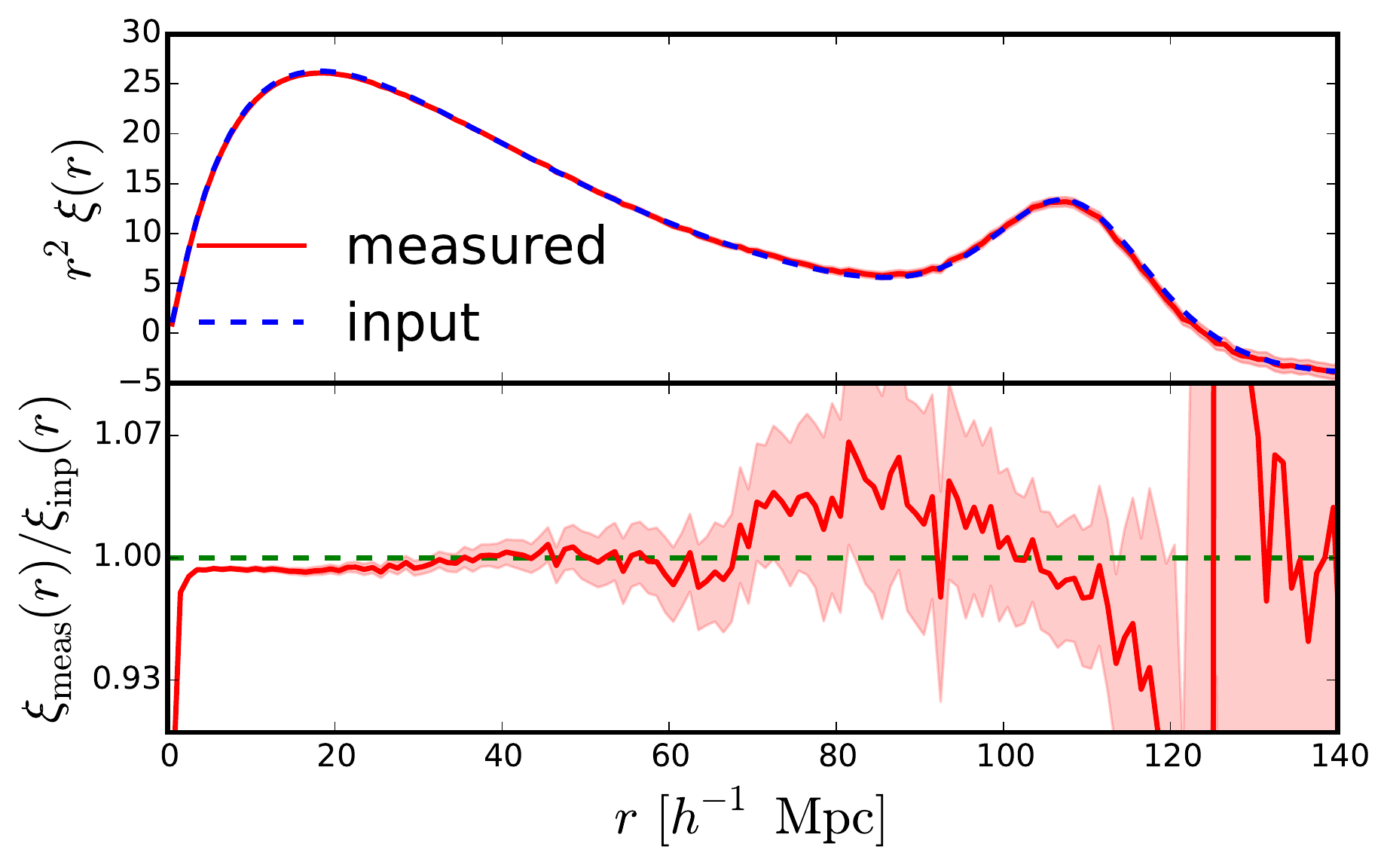}
 \caption{
 Same as \reffig{pk_real} but for the correlation function. We show
 $r^2\xi(r)$. The
 spikes at $r\approx 120-130\hMpc$ in the bottom panel corresponds to
 the zero crossing of the correlation function so a small deviation in
 the mean of the measured correlation function leads to a large ratio. We use a bin size of $1\, h^{-1}$ Mpc. 
 }
\label{fig:xi_real}
\end{figure}
%%%%%%%%%%%%%%%%%%%%%%%%%%%%%%%%%%%%%%%%%%%%%%%%%%%%%%%%%%%%%%%%%%%%%%%%

We measure the galaxy two-point correlation function by using the Landy-Szalay 
estimator \cite{landy:1993}
\begin{equation}
 \xi(\vr) = \frac{DD(\vr)-2DR(\vr) + RR(\vr)}{RR(\vr)} \,,
 \label{eq:ls_est}
\end{equation}
where $DD(\vr)$, $DR(\vr)$, and $RR(\vr)$ are the number of galaxy-galaxy,
galaxy-random, and random-random pairs, respectively.
The Landy-Szalay estimator cancels the leading order uncertainties in 
estimating the mean number density.
\refFig{xi_real} shows the average of the measured 
correlation function (top) and the ratio to the input (bottom).
We also find an excellent agreement between the measured and input correlation
functions over a wide range of scales.

%%%%%%%%%%%%%%%%%%%%%%%%%%%%%%%%%%%%%%%%%%%%%%%%%%%%%%%%%%%%%%%%%%%%%%%%
\subsubsection{Cross-Correlation Coefficient}
\label{sec:real_space_cxi}
%%%%%%%%%%%%%%%%%%%%%%%%%%%%%%%%%%%%%%%%%%%%%%%%%%%%%%%%%%%%%%%%%%%%%%%%

\begin{figure}[t]
	\centering
	\includegraphics[width=0.8\textwidth, bb=0 0 533 333]{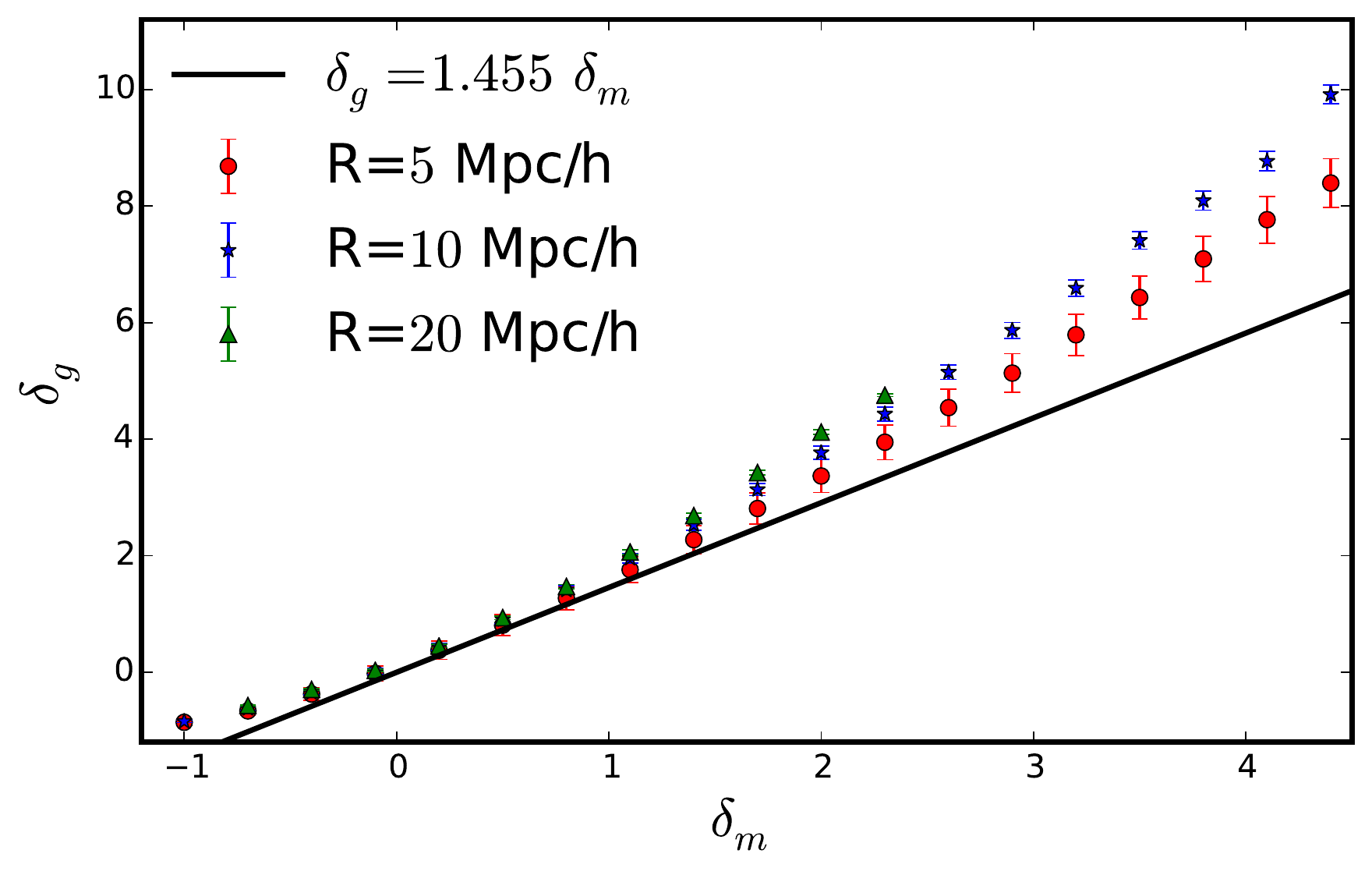}
	\caption{{Galaxy overdensity $\delta_g$ as a function of matter overdensity 
	$\delta_m$ smoothed over different scales - $5\, h^{-1}\rm{Mpc}$ (red circles), $10\, 
	h^{-1}\rm{Mpc}$ (blue stars) and $20\, h^{-1}\rm{Mpc}$ (green triangles). The error bars 
	show the error on the mean of the galaxy overdensity, averaged over all the cells in our 
	50 mocks.}}
	\label{fig:dg_dm}
\end{figure}

\begin{figure}[t]
\centering
\includegraphics[width=0.8\textwidth, bb = 0 0 517 332]{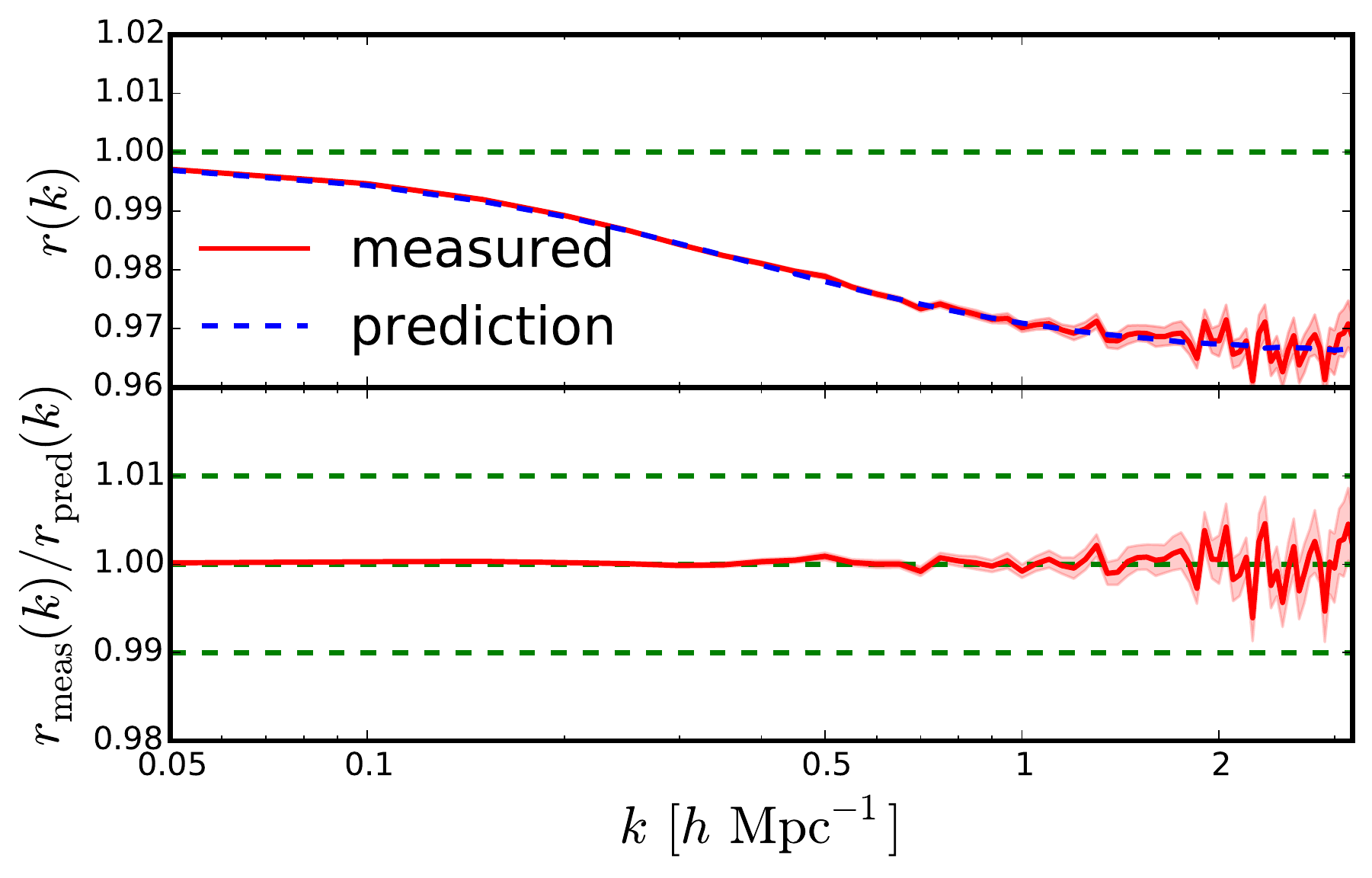}
\caption{(Top) Measured (red solid) and predicted (blue dashed)
 cross-correlation coefficient as a function of the wavenumber. The red
 band shows the error on the mean measured from 50 log-normal mock
 catalogs. (Bottom) Ratio of the measured to the predicted
 cross-correlation coefficients.}
\label{fig:cpk}
\end{figure}

We next examine the cross-correlation coefficient
\be
 r(k)=\frac{P_{gm}(k)}{\sqrt{P_{gg}(k)P_{mm}(k)}} \,,
\label{eq:cpk}
\ee
between matter and galaxy density contrasts in real space. Here, 
$P_{gm}(k)$ denotes the cross power spectrum of galaxy and matter. 
In \reffig{cpk}, the red solid line in the top panel shows the measured 
cross-correlation coefficient $r(k)$ from the log-normal mock catalogs. 
The measured cross-correlation coefficient approaches unity on large
scales, but decreases on small scales with high significance,
despite the fact that we have imposed a linear bias relation
between the galaxy and matter power spectra, $P_{gg}(k)=b^2 P_{mm}(k)$,
and that the random realisations of $G_g$ and $G_m$ have been
drawn from an identical random seed. This result is a generic feature of
log-normal fields that the proportionality relation in power spectra
does not guarantee the proportionality of the fields \cite{Xavier:2016elr}.
{We can also see this more clearly in a plot of smoothed galaxy and 
matter density fields, as in \reffig{dg_dm}. The smoothed galaxy overdensity deviates 
significantly from the linearly biased one with $b=1.455$ for all smoothing 
scales that we chose. In fact, the bias is seen to not even be linear. This is a consequence 
of the non-linear transformation between the Gaussian and log-normal fields. }  

%%%%%%%%%%%%%%%%%%%%%%%%%%%%%%%%%%%%%%%%%%%%%%%%%%%%%%%%%%%%%%%%%%%%%%%%%%%%%

%%%%%%%%%%%%%%%%%%%%%%%%%%%%%%%%%%%%%%%%%%%%%%%%%%%%%%%%%%%%%%%%%%%%%%%%%%%%%

%%%%%%%%%%%%%%%%%%%%%%%%%%%%%%%%%%%%%%%%%%%%%%%%%%%%%%%%%%%%%%%%%%%%%%%%
%%%%%%%%%%%%%%%%%%%%%%%%%%%%%%%%%%%%%%%%%%%%%%%%%%%%%%%%%%%%%%%%%%%%%%%%

Because, in our mock, the Gaussian (log-transformed) fields of 
galaxy and matter have the same random numbers
($\theta_r+i\theta_i$ in \refeq{Gk}), they are related to each other in 
every Fourier cell as 
\be
 \frac{G_g(\vk)}{G_m(\vk)}=\sqrt{\frac{P^G_{gg}(k)}{P^G_{mm}(k)}} \,,
\ee
where $P^G_{mm}(k)$ and $P^G_{gg}(k)$ are the power spectra of $G_m$ and
$G_g$, respectively. Indeed, it is the cross-correlation coefficient
between $G_m$ and $G_g$ that is equal to unity. On the other hand, the
galaxy and matter density fields $\delta_g(\vx)$ and $\delta_m(\vx)$ are
exponentially related to $G_g(\vx)$ and $G_m(\vx)$. That is,
$\delta_g(\vk)$ and $\delta_m(\vk)$ are not linearly related to each
other so that the cross-correlation coefficient must deviate from one
\cite{Xavier:2016elr}. On large scales where the correlation functions
are small, the cross-correlation approaches unity because $\delta$'s are
approximately the same as $G$'s.

We can compute $r(k)$ analytically for log-normal density fields.
Specifically, using the one-dimensional Fourier transform we first compute
the galaxy-matter cross spectrum as
\be
 P_{gm}(k)=\int dr~r^2 \xi_{gm}(r) j_0(kr)
 = \int dr~r^2 \left[ e^{\xi^G_{gm}(r)} -1 \right] j_0(kr) \,,
\label{eq:sph_ft}
\ee
where $j_0(x)$ is the spherical Bessel function of the zeroth order, and 
$\xi_{gm}({r})=\langle\delta_g(\vx+\vr)\delta_m(\vx)\rangle$ and
$\xi^G_{gm}(r)=\langle G_g(\vx+\vr)G_m(\vx)\rangle$. The relation
between $\xi_{gm}({r})$ and $\xi^G_{gm}(r)$ is analogous to \refeq{xiG
and xi}. We then compute $\xi^G_{gm}(r)$ as
\ba
 \xi^G_{gm}(r) \:&
 = \int \frac{d^3k_1}{(2\pi)^3}\frac{d^3k_2}{(2\pi)^3} \langle {G_g(\vk_1)}{G^{*}_m(\vk_2)}\rangle
      e^{i[\vk_1\cdot(\vx+\vr)-\vk_2\cdot \vx]} \vs
 \:& = \int \frac{d^3k_1}{(2\pi)^3}\frac{d^3k_2}{(2\pi)^3} \sqrt{\frac{P^G_{gg}(k_1)}{P^G_{mm}(k_1)}}
      \langle{G_m(\vk_1)}{G^{*}_m(\vk_2)}\rangle e^{i[(\vk_1-\vk_2)\cdot\vx+\vk_1\cdot\vr]} \vs
 \:& = \int \frac{dk}{2\pi^2} k^2\sqrt{P^G_{gg}(k)P^G_{mm}(k)} j_0(kr) \,.
\label{eq:pGgm}
\ea
Combining \refeqs{sph_ft}{pGgm}, $r(k)$ can be evaluated.
The blue dashed line in the top panel of \reffig{cpk} shows the prediction,
whereas the bottom panel shows the ratio between the measurement and the
prediction. We find an excellent agreement between the measurement and the
prediction.

%%%%%%%%%%%%%%%%%%%%%%%%%%%%%%%%%%%%%%%%%%%%%%%%%%%%%%%%%%%%%%%%%%%%%%%%
\subsection{Redshift-space density statistics}
\label{sec:zspace_stat}
%%%%%%%%%%%%%%%%%%%%%%%%%%%%%%%%%%%%%%%%%%%%%%%%%%%%%%%%%%%%%%%%%%%%%%%%
We now present the measurements of the two-point statistics in redshift space. 
We obtain the redshift-space mock catalogs from the real-space ones by 
mapping real-space positions of galaxies to redshift-space positions by
\refeq{r_to_s}.
%\begin{equation}
%\vecs = \vx + \frac{v_z(\vx)}{\mathcal{H}}\hat{z} \,.
%\end{equation}
We use the periodic boundary condition along the $z$-direction for
galaxies that move out of the box by this mapping.
Measurements of the power spectrum or correlation
function for these redshift-space catalogs proceed in the same
manner as in real space.

As described in \refsec{review}, when linearising the Jacobian, the 
redshift-space power spectrum is given by
\begin{equation}
\label{eq:kaiser}
 P^s_{gg}(k, \mu_k) = P_{gg}(k)+2\mu_k^2fP_{gm}(k)+\mu_k^4f^2P_{mm}(k).
\end{equation}
When using linear theory (that we shall call ``Kaiser''), we  
relate the galaxy-galaxy power spectrum and galaxy-matter power spectrum
to the matter-matter power spectrum as $P_{gg}(k) = b^2 P_{mm}(k)$
and $P_{gm}(k)=bP_{mm}(k)$.
We stress, however, that the galaxy-matter cross power spectrum $P_{gm}(k)$ 
is not equal to $bP_{mm}(k)$ for the log-normal density fields, as shown 
in \refsec{real_space_cxi}. Therefore, in order to highlight the effect
from non-linearity in the Jacobian, we calculate the redshift-space galaxy
power spectrum with \refeq{kaiser} but use the cross power spectrum in 
\refsec{real_space_cxi} (that we call ``linear Jacobian''). 

\begin{figure}[t]
\centering
\includegraphics[width=0.8\textwidth, bb=0 0 522 327]{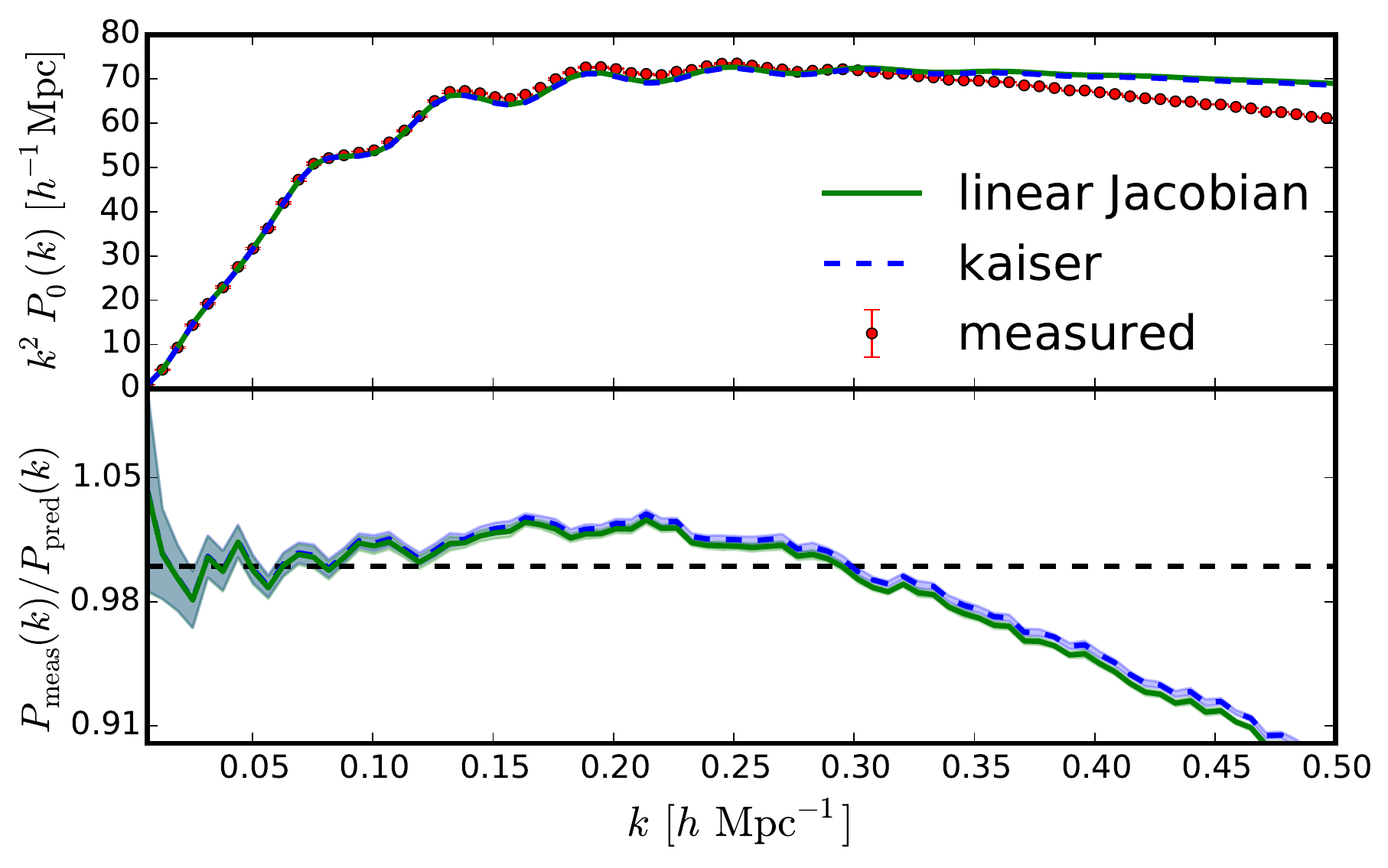}
\caption{{(Top) Monopole redshift-space power spectrum. The red points
 show the measurement averaged over 50 log-normal mock
 catalogs. The blue dashed line shows the Kaiser prediction while the
 green solid line shows the linear Jacobian prediction. We show
 $k^2P_0(k)$ to enhance differences at large $k$. (Bottom) Ratio of the
 measured monopole power to the Kaiser and linear Jacobian
 predictions. The band shows the error on the mean estimated from 50
 realisations. We find a sub-1\% agreement on scales $k \apprle 0.1
 \ihMpc$.}
}
\label{fig:kaiser_pk0}
\end{figure}

\refFigs{kaiser_pk0}{kaiser_pk2} show the measured monopole and 
quadrupole power spectra, compared with the Kaiser (with $P_{gm}(k)=bP_{mm}(k)$)
and the linear Jacobian (with measured $P_{gm}(k)$) predictions. 
On large scales (small $k$), we find a good agreement for all three cases 
as expected. On small scales, the linear Jacobian calculation deviates from the 
Kaiser value due to the non-unity cross-correlation between the matter and
galaxy (\refsec{real_space_cxi}). The deviation is smaller than what we find 
in \reffig{cpk} because $f/b = 0.616$ is less than unity.

%%%%%%%%%%%%%%%%%%%%%%%%%%%%%%%%%%%%%%%%%%%%%%%%%%%%%%%%%%%%%%%%%%%%%%%%
%%%%%%%%%%%%%%%%%%%%%%%%%%%%%%%%%%%%%%%%%%%%%%%%%%%%%%%%%%%%%%%%%%%%%%%%

%%%%%%%%%%%%%%%%%%%%%%%%%%%%%%%%%%%%%%%%%%%%%%%%%%%%%%%%%%%%%%%%%%%%%%%%
\begin{figure}[t]
\centering
\includegraphics[width=0.8\textwidth, bb=0 0 522 327]{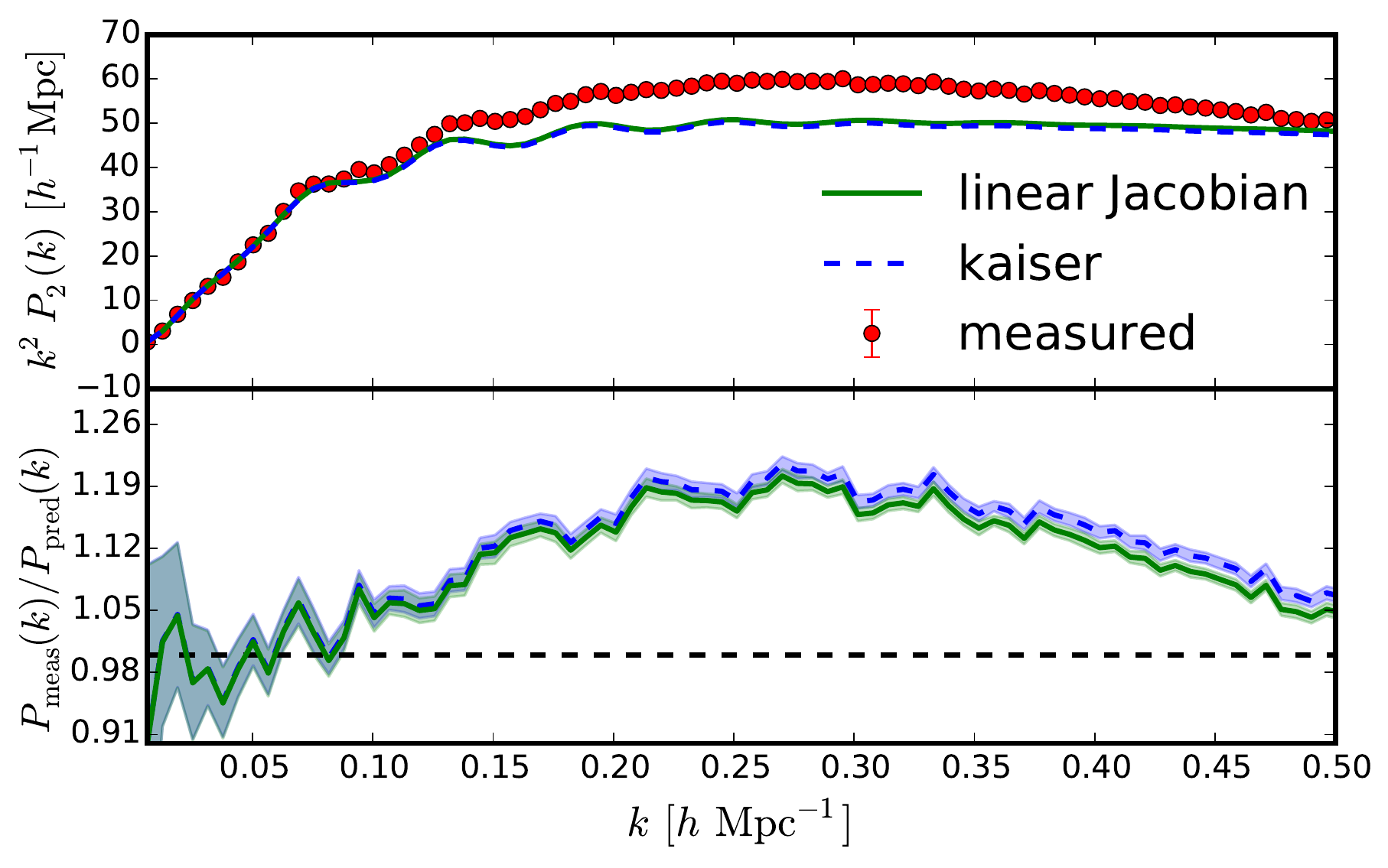}
\caption{Same as \reffig{kaiser_pk0}, but for the quadrupole power spectrum.
}
\label{fig:kaiser_pk2}
\end{figure}
%%%%%%%%%%%%%%%%%%%%%%%%%%%%%%%%%%%%%%%%%%%%%%%%%%%%%%%%%%%%%%%%%%%%%%%%

Both the Kaiser and linear Jacobian calculations fail to model the
measured monopole and quadrupole power spectra on small scales. As we
use the real-space power spectrum used to generate the log-normal
catalog in each term in \refeq{kaiser}, any discrepancy that we find in 
\reffigs{kaiser_pk0}{kaiser_pk2} is due to non-linearity in the mapping 
between real and redshift space.
That is, when densities and velocities become large, the Jacobian cannot 
be linearised, and so the linear Jacobian approximation is no longer valid. 
For example, the measured monopole power spectrum in \reffig{kaiser_pk0}
is smaller than the linear Jacobian calculation on small scales 
($k>0.3\,h\,{\rm Mpc}^{-1}$). This behaviour is qualitatively similar to the
FoG damping effect, which is usually attributed to RSD of random motion
within bigger halos. In our log-normal mock catalog, however, the
damping cannot be the FoG effect because we do not include any random
component when generating peculiar velocities. Rather, the power
suppression we see here originates solely from non-linearity in the Jacobian
of real-to-redshift mapping due to {\it coherent} peculiar velocity
fields given by the continuity equation.

Mathematically, combining \refeqs{dgs_dgJ}{jacobian} yields the
the non-linear mapping between the real- and redshift-space density contrasts
\begin{equation}
 1+\delta^s_g(\vecs) = \frac {1+\delta_g(\vx)}{\left| 1+\frac{1}{\mathcal{H}}
 \frac{\partial v_z(\vx)}{\partial z}\right| } \,,
\end{equation}
which turns to, in Fourier space \cite{Taruya:2010mx},
\be
\label{eq:deltas_k}
 \delta^s_g(\vk) = \int d^3x\, e^{-i\vk\cdot\vx}
 \left[ \delta_g(\vx)-\frac{1}{\mathcal{H}}\frac{\partial v_z(\vx)}{\partial z}\right]
 e^{ik_zv_z(\vx)/\mathcal{H}} \,.
\ee
As long as velocities are small, i.e. $k_zv_z \ll 1$, the exponential factor in
\refeq{deltas_k} can be approximated to unity, leading to the linear Jacobian 
formula. However, on small scales, this approximation breaks down, and the 
exponential factor leads to the non-linear Jacobian effects  
\cite{Taruya:2010mx}. 

\begin{figure}[t]
\centering
\includegraphics[width=0.8\textwidth, bb=0 0 510 327]{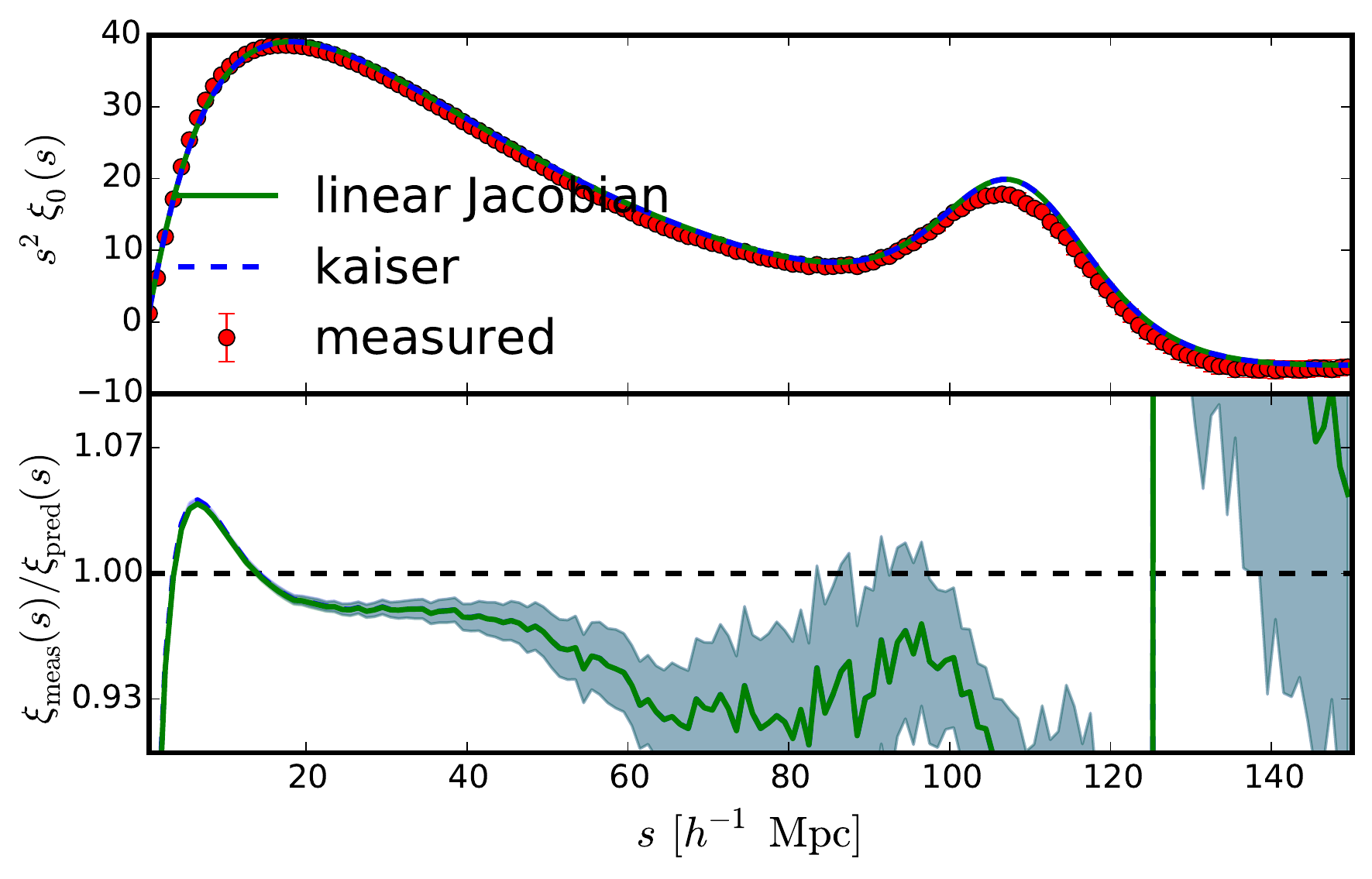}
\caption{(Top) Monopole redshift-space two-point correlation
 function. The meaning of the lines is the same as in
 \reffig{kaiser_pk0}. We show $s^2\xi_0(s)$ to enhance differences on
 large separations. (Bottom) Ratio of the measured monopole correlation
 function to the Kaiser and linear Jacobian predictions. {The Kaiser 
 and linear Jacobian predictions lie almost on top of each other, and so 
 are hard to distinguish.}
}
\label{fig:kaiser_xi0}
\end{figure}

We show the configuration-space two-point correlation function in 
\reffigs{kaiser_xi0}{kaiser_xi2}, and compare them with the linearised 
Jacobian prediction:
\be
\delta_g^s(\vecs) = \delta_g(\vx) - \frac{1}{\cH} \frac{\partial}{\partial z}v_z(\vx)\,.
\ee
Combining with the linear continuity equation that we use to generate the 
velocity field,
\be
\vv(\vx) = -\cH f\nabla
\left[\nabla^{-2}\delta_m(\vx)\right],
\ee
we find the configuration space expression for the redshift-space density 
contrast:
\be
\delta_g^s(\vecs) = \delta_g + f \left(\frac{\partial}{\partial z}\right)^2
\left[\nabla^{-2}\delta_m(\vx)\right],
\ee
from which we calculate
\ba \label{eq:kaiser_xi}
\xi^s_{gg}(s,\mu)\:&=
\left[\xi_{gg}(s)+\frac23f\xi_{gm}(s) +\frac15f^2\xi_{mm}(s) \right] \vs
\:&-\left[\frac43 f\xi_{gm,2}(s)+\frac47f^2\xi_{mm,2}(s) \right]{\cal L}_2(\mu)
+\frac{8}{35}f^2\xi_{mm,4}(s){\cal L}_4(\mu).
\ea
Just like the case for the power spectrum, the expression reduces to the 
linear Kaiser prediction (\refeq{def_xils}) when we set $\xi_{gm}(r) =
b\xi_{mm}(r)$, but we must take into account non-unity cross-correlation
function for the log-normal catalog.

While the Kaiser and linear Jacobian models are reproduced in the power
spectrum at $k\lesssim 0.1~h~{\rm Mpc}^{-1}$, they are not well
reproduced in the correlation functions at all separations. This
indicates that the correlation functions at large separations are
sensitive to non-linearity in the Jacobian.

%%%%%%%%%%%%%%%%%%%%%%%%%%%%%%%%%%%%%%%%%%%%%%%%%%%%%%%%%%%%%%%%%%%%%%%%
%%%%%%%%%%%%%%%%%%%%%%%%%%%%%%%%%%%%%%%%%%%%%%%%%%%%%%%%%%%%%%%%%%%%%%%%

%%%%%%%%%%%%%%%%%%%%%%%%%%%%%%%%%%%%%%%%%%%%%%%%%%%%%%%%%%%%%%%%%%%%%%%%
\begin{figure}[t]
\centering
\includegraphics[width=0.8\textwidth, bb=0 0 510 327]{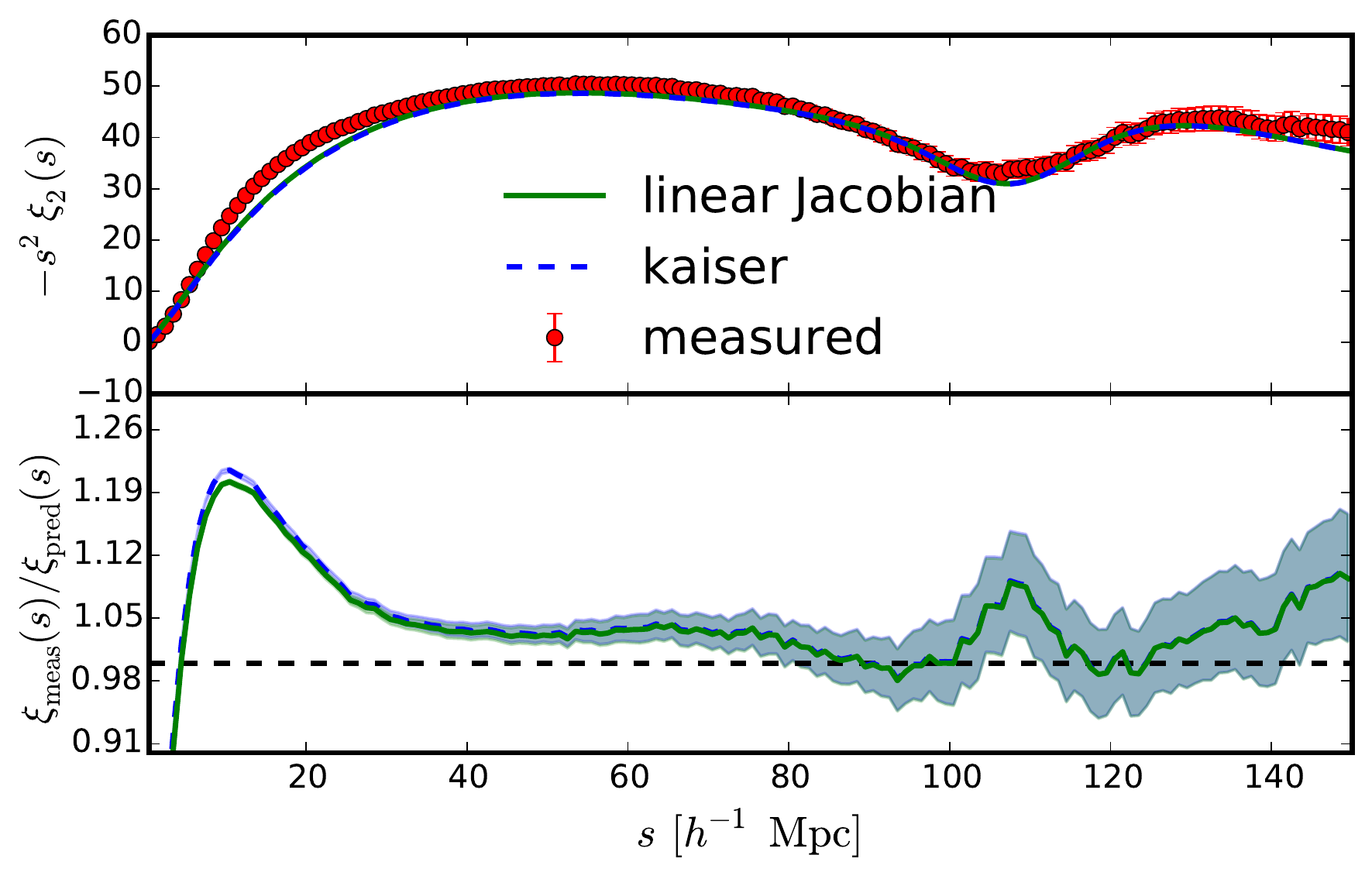}
\caption{Same as \reffig{kaiser_xi0}, but for the quadrupole two-point
correlation function. We show $-s^2\xi_2(s)$.
}
\label{fig:kaiser_xi2}
\end{figure}
%%%%%%%%%%%%%%%%%%%%%%%%%%%%%%%%%%%%%%%%%%%%%%%%%%%%%%%%%%%%%%%%%%%%%%%%

%%%%%%%%%%%%%%%%%%%%%%%%%%%%%%%%%%%%%%%%%%%%%%%%%%%%%%%%%%%%%%%%%%%%%%%%
\subsection{Pairwise Line-of-Sight Velocity PDFs}
\label{sec:real_space_pdf}
%%%%%%%%%%%%%%%%%%%%%%%%%%%%%%%%%%%%%%%%%%%%%%%%%%%%%%%%%%%%%%%%%%%%%%%%

How do we incorporate non-linearity in the Jacobian into the model? 
As discussed in \refsec{review}, the pairwise line-of-sight velocity PDF fully
describes the mapping from the real-space two-point correlation function to 
the redshift-space one. 
%%%%%%%%%%%%%%%%%%%%%%%%%%%%%%%%%%%%%%%%%%%%%%%%%%%%%%%%%%%%%%%%%%%%%%%%
\begin{figure}[t]
\centering
\includegraphics[width=1\textwidth, bb=0 0 493 305]{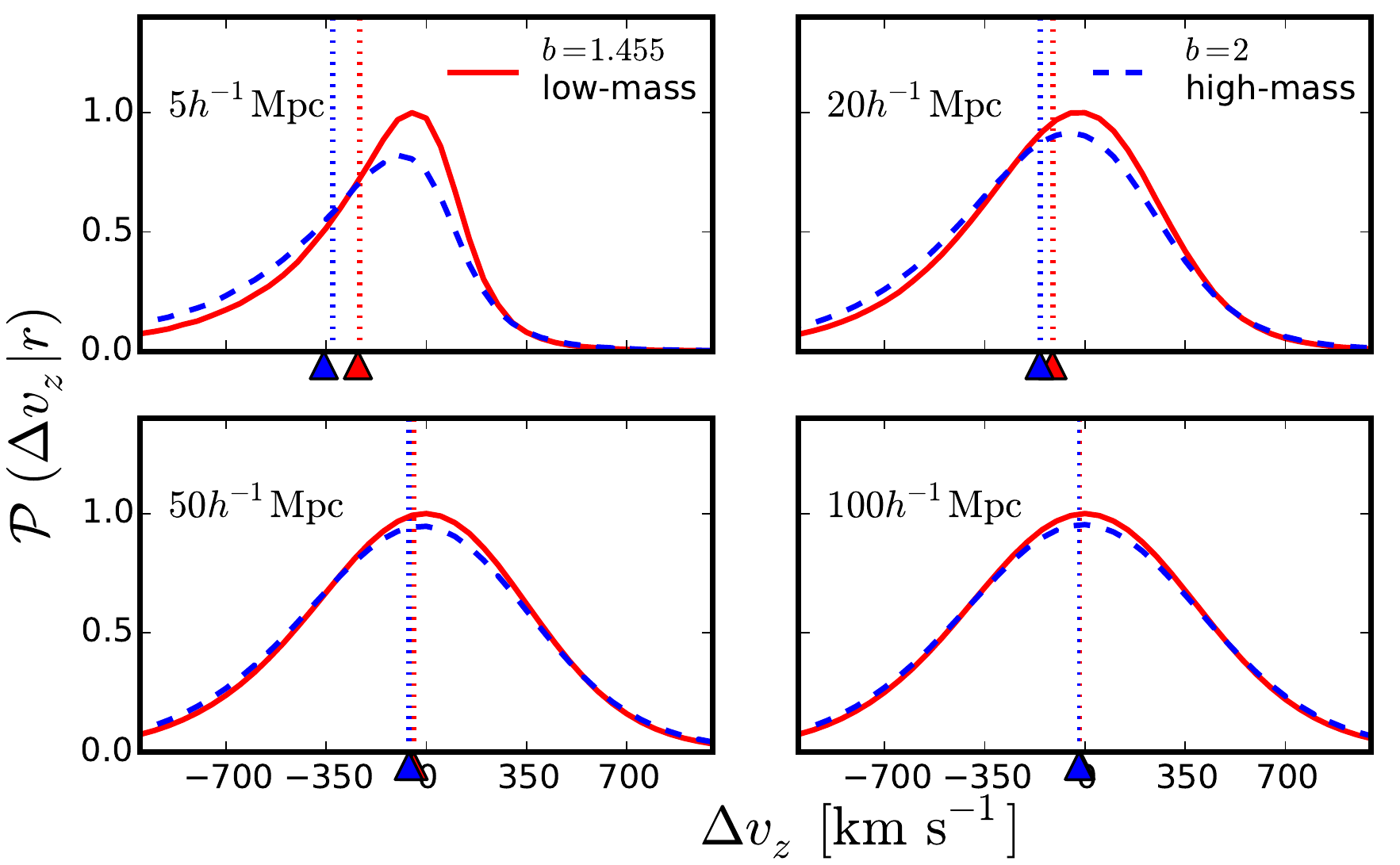}
 \caption{Same as \reffig{pdf_nbody}, but for PDFs averaged over 50
 log-normal mock catalogs. The simulation volume is $(1000\hMpc)^3$ and
 the output is at $z=1.3$. The biases of low- and high-mass catalogs are
 $b=1.455$ (red) and 2 (blue), respectively. The triangles on the
 horizontal axes show the calculations from \refeq{mgf_0}.}
\label{fig:pdf_ln}
\end{figure}
%%%%%%%%%%%%%%%%%%%%%%%%%%%%%%%%%%%%%%%%%%%%%%%%%%%%%%%%%%%%%%%%%%%%%%%%

We show in \reffig{pdf_ln} the PDFs of pairwise line-of-sight 
velocity averaged over 50 log-normal mock catalogs, for four different 
separations between galaxy pairs ({\it From top left to bottom right}, 
$5.25$, $20.25$, $50.25$, and $100.25\hMpc$) along the line-of-sight
direction ($0.99\leq\mu\leq1$). We show the pairwise line-of-sight
velocity PDFs along the line-of-sight direction because the relative
peculiar velocities are at their maximum (i.e., no perpendicular
component). For the streaming model, we need the relative 
velocity PDFs for galaxy pairs along all directions.

We show the measured pairwise line-of-sight velocity PDFs 
for two different linear biases $b=1.455$ (low-mass) and 2 (high-mass) as,
respectively, the blue dashed lines and the red solid lines.
Note that we use the same phases (that is, the same sequence of random numbers)
for generating velocity fields for both cases; thus, galaxies in the same cell have identical velocities regardless of the assumed bias parameter.
Overall, we find that the pairwise line-of-sight velocity PDFs from our
log-normal mock catalogs capture qualitative features that we have seen
in N-body simulations (\reffig{pdf_nbody}).
Namely, both PDFs have negative mean velocity and negative skewness for 
smaller separations (top panels of \reffig{pdf_ln} and \reffig{pdf_nbody})
and approach a symmetric PDF for larger separations.
Also, the tendency is more obvious for high-mass (high-bias) galaxies.
The log-normal catalogs show larger velocity dispersion than N-body
simulations especially at small separations.

Our results show that the coherent irrotational velocity given by the
linearised continuity equation can explain a part of the non-linear
features in the pairwise line-of-sight velocity PDF. We stress, again,
that we do not include any random velocities. 
Also note that we have
assigned the same velocity to all galaxies in the same cell, and the
velocity field is exactly the same for the high-mass and low-mass
samples. Nevertheless, the pairwise line-of-sight velocity PDF for galaxies with 
different biases are still different due to 
pair weighting: velocities of galaxies with high (low) bias are weighted higher (lower) density regions.

While successful at a qualitative level, the PDFs from log-normal
catalogs and those from N-body simulations are different in detail. For
example, the velocity field in our log-normal mock catalogs only
reflects non-Gaussianity of the log-normal density fields, which is not
the same as that in N-body simulations where the complete non-linear 
gravitational evolution is encoded.

Strong non-Gaussianity in the pairwise line-of-sight velocity PDF makes
it challenging to analytically compute its full moments, even if the
velocity field is assumed to follow the linear continuity
equation. Nevertheless, we can still compute the mean pairwise line-of-sight 
velocity (that is, the first moment) as follows. 
Using the linearised continuity equation (\refeq{cont_eq_k}), the parallel 
component of relative velocity for galaxies separated by $r$ is given by
\begin{equation}
v_{1z}-v_{2z} = i\mathcal{H}f\int \frac{d^3k}{(2\pi)^3} \frac{k_z}{k^2}\delta_m(\vk)\left(e^{i\vk\cdot \vx_1}-e^{i\vk\cdot \vx_2}\right) \,.
\end{equation}
Because the parallel relative velocity is the exponent in the velocity 
generating function (\refeq{mgf_pdf}), the mean pairwise line-of-sight 
velocity can be calculated from 
\begin{align}
 \langle \Delta v_{z} \rangle = 
\cH\left.\frac{\partial M(\lambda,\vr)}{\partial\lambda}\right|_{\lambda=0}\:&=
 \frac{\left\langle\left( v_{1z}-v_{2z}\right) \left[ 1+\delta_{g1}\right] \left[ 1+\delta_{g2}\right]\right\rangle}{1+\xi_{gg}(r)} \vs
 \:&=i\mathcal{H}f\frac{\int d^3x' q(\vx'_{1}, \vx'_{2})\left\langle \left[ 1+\delta_{g1}\right]
 \left[ 1+\delta_{g2}\right]\delta_{m3}\right\rangle}{1+\xi_{gg}(r)} \,,
\label{eq:mgf_0}
\end{align}
where $\vx'_i \equiv \vx_i-\vx'$, $\delta_{m3} = \delta_m(\vx')$, and
\begin{equation}
 q(\vx'_1,\vx'_2) \equiv \int\frac{d^3k}{(2\pi)^3} \frac{k_z}{k^2}
\left(e^{i\vk\cdot\vx'_1}-e^{i\vk\cdot\vx'_2}\right) \,.
\end{equation}
The triangles on the horizontal axes in \reffig{pdf_ln} show the
predictions computed from \refeq{mgf_0}. We find that they are in an excellent agreement with the measurements. As shown in \refeq{mgf_0} the mean of the pairwise
line-of-sight velocity depends on the integral of the three-point function of
the log-normal fields. We present the details of this calculation in
\refapp{mean_pairwise_v}.

Likewise, in order to compute the $n$th-order moments of the pairwise 
line-of-sight velocity we need to integrate over $(n+2)$-point correlation 
functions of the log-normal fields. Thus, it is impractical to compute all the 
moments of the pairwise line-of-sight velocity PDF, even though the statistics 
of log-normal fields are known. 
On the other hand, should we assume a Gaussian PDF for the pairwise 
line-of-sight velocity distribution, all the higher-order moments but
the first two would be ignored; thus, we miss the non-Gaussian effects coming 
from non-linear evolution of the Universe as well as the pair weighting effect.

%%%%%%%%%%%%%%%%%%%%%%%%%%%%%%%%%%%%%%%%%%%%%%%%%%%%%%%%%%%%%%%%%%%%%%%%
\subsection{Recovery of Kaiser limit}
\label{sec:kaiser_limit}
%%%%%%%%%%%%%%%%%%%%%%%%%%%%%%%%%%%%%%%%%%%%%%%%%%%%%%%%%%%%%%%%%%%%%%%%

\begin{figure}[t]
	\centering
	\includegraphics[width=1\textwidth, bb=0 0 537 332]{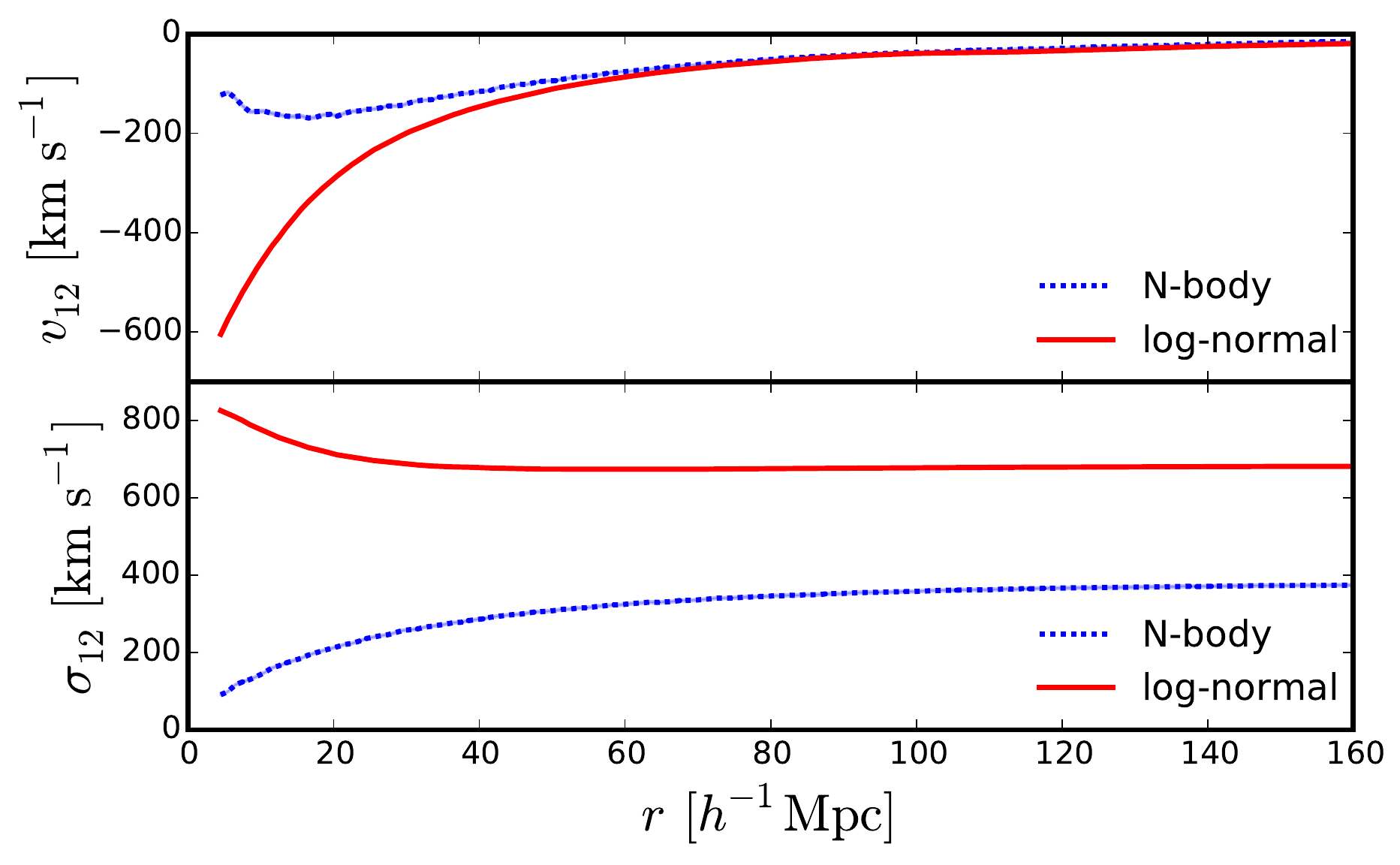}
	\caption{{Mean (top) and dispersion (bottom) of pairwise velocity from log-normal mocks (red) and N-body simulations (blue) as a function of pair separation $r$, for pairs along the line-of-sight, $0.99 \leq \mu \leq 1.0$. The linear bias for log-normal mocks was chosen to be $b=1.8$ which is the average linear bias for the halos in the mass range $5 \times 10^{13} - 6.5 \times 10^{13}$ $h^{-1}\,M_{\odot}$ shown above.}}
	\label{fig:v_moments}
\end{figure}

Even though the pairwise velocity PDF in our log-normal mocks does not precisely reproduce the velocity PDF from N-body simulations (see for example, \reffig{v_moments}, for a comparison of the first two moments of the pairwise velocities as a function of separation), the redshift space power spectrum from both matches the Kaiser prediction on large scales ($k \lesssim 0.1\, h$ Mpc$^{-1}$). To understand why this happens, we now consider the redshift space two-point correlation function in the large-scale limit, i.e. separations such that only the linear order terms contribute.  

%%%%%%%%%%%%%%%%%%%%%%%%%%%%%%%%%%%%%%%%%%%%%%%%%%%%%%%%%%%%%%%%%%%%%%%%
%%%%%%%%%%%%%%%%%%%%%%%%%%%%%%%%%%%%%%%%%%%%%%%%%%%%%%%%%%%%%%%%%%%%%%%%

The large-scale limit of the redshift space two-point correlation function (to linear order) is given as \cite{Fisher:1994ks}
\begin{equation}\label{eq:large_r_xis}
\xi^s_{gg}(s_{\para}, s_{\perp}) = \xi_{gg}(s)-\frac{\dd}{ \dd r_{\para}}\Big[v_{12}(r)\frac{r_{\para}}{r}\Big]\Big| _{r_{\para}=s_{\para}}+\frac{1}{2}\frac{ \dd^2}{ \dd r_{\para}^2}\Big[\sigma^2_{12}(r, \mu)\Big]\Big|_{r_{\para}=s_{\para}} \,,
%+\frac{\sigma^2_{12}|_{\infty}}{2}\frac{ \dd^2}{ \dd y^2}\Big[\xi(r)\Big]\Big|_{y=s_{\para}}
\end{equation}
where $s^2 \equiv s^2_{\para}+s^2_{\perp}$, $r^2 \equiv s^2_{\perp}+r_{\para}^2$, $v_{12}(r)$ is the mean of the radial pairwise velocity (which is the relative velocity projected along the line joining the pair of particles; we call the remaining component tangential pairwise velocity), and $\sigma^2_{12}(r, \mu)$ is the variance of the line-of-sight pairwise velocities, with $\mu \equiv r_{\para}/r$. 
%Taking the monopole of the redshift-space correlation function as in \refeq{large_r_xis}, one can show that the first two terms reduce to the standard Kaiser factor (\refeq{kaiser_xi}) but the last term contributes a 

In linear theory, the mean and variance can be calculated as
\begin{align}
v_{12}(r) &= -\frac{\mathcal{H}fb}{\pi^2}\int dk k P_{\rm mm}(k) j_1(kr) \,, \\
\sigma^2_{12}(r,\mu) &= 2[\sigma^2_v-\mu^2\Psi_{\para}-(1-\mu^2)\Psi_{\perp}] \,, \\
\Psi_{\para}(r) &= \frac{\mathcal{H}^2f^2}{2\pi^2}\int dk P_{\rm mm}(k) \Bigg[j_0(kr)-\frac{2j_1(kr)}{kr}\Bigg] \,, \\
\Psi_{\perp}(r) &= \frac{\mathcal{H}^2f^2}{2\pi^2}\int dk P_{\rm mm}(k) \frac{j_1(kr)}{kr} \,,
\end{align}
where $\sigma^2_v \equiv \left\langle \vv(\vx) \cdot \vv(\vx)\right\rangle/3 $ is the one-dimensional velocity variance, $P_{\rm mm}(k)$ is the matter power spectrum in real space, $\Psi_{\para}$ is the variance of the radial pairwise velocities, $\Psi_{\perp}$ is the variance of the tangential pairwise velocities, and $j_n$ denotes the spherical Bessel function of the $n^{\rm th}$ order. As shown in refs.~\cite{Fisher:1994ks,reid:2011},
\begin{align}
-\frac{\dd}{\dd r_{\para}}\Big[v_{12}(r)\frac{r_{\para}}{r}\Big]\Big| _{r_{\para}=s_{\para}}&=\frac{v_{12}(r)}{r}(\mu^2-1)-v'_{12}(r)\mu^2\, ,\\
\nonumber\frac{1}{2}\frac{\dd^2}{\dd r_{\para}^2}\Big[\sigma^2_{12}(r, \mu)\Big]\Big|_{r_{\para}=s_{\para}}&=(2-10\mu^2+8\mu^4)\frac{\Psi_{\perp}-\Psi_{\para}}{r^2}+(5\mu^4-5\mu^2)\frac{\Psi'_{\para}}{r}\\
&+(-1+6\mu^2-5\mu^4)\frac{\Psi'_{\perp}}{r}-\mu^4\Psi''_{\para}+(\mu^4-\mu^2)\Psi''_{\perp}\, ,
\end{align}
so that the Kaiser limit of the redshift space two-point correlation function does not depend on any constant or the isotropic dispersion of the pairwise line-of-sight velocities. In this section $'$ denotes $d/dr$.

%%%%%%%%%%%%%%%%%%%%%%%%%%%%%%%%%%%%%%%%%%%%%%%%%%%%%%%%%%%%%%%%%%%%%%%%
\begin{figure}[t]
	\centering
	\includegraphics[width=1\textwidth, bb=0 0 508 314]{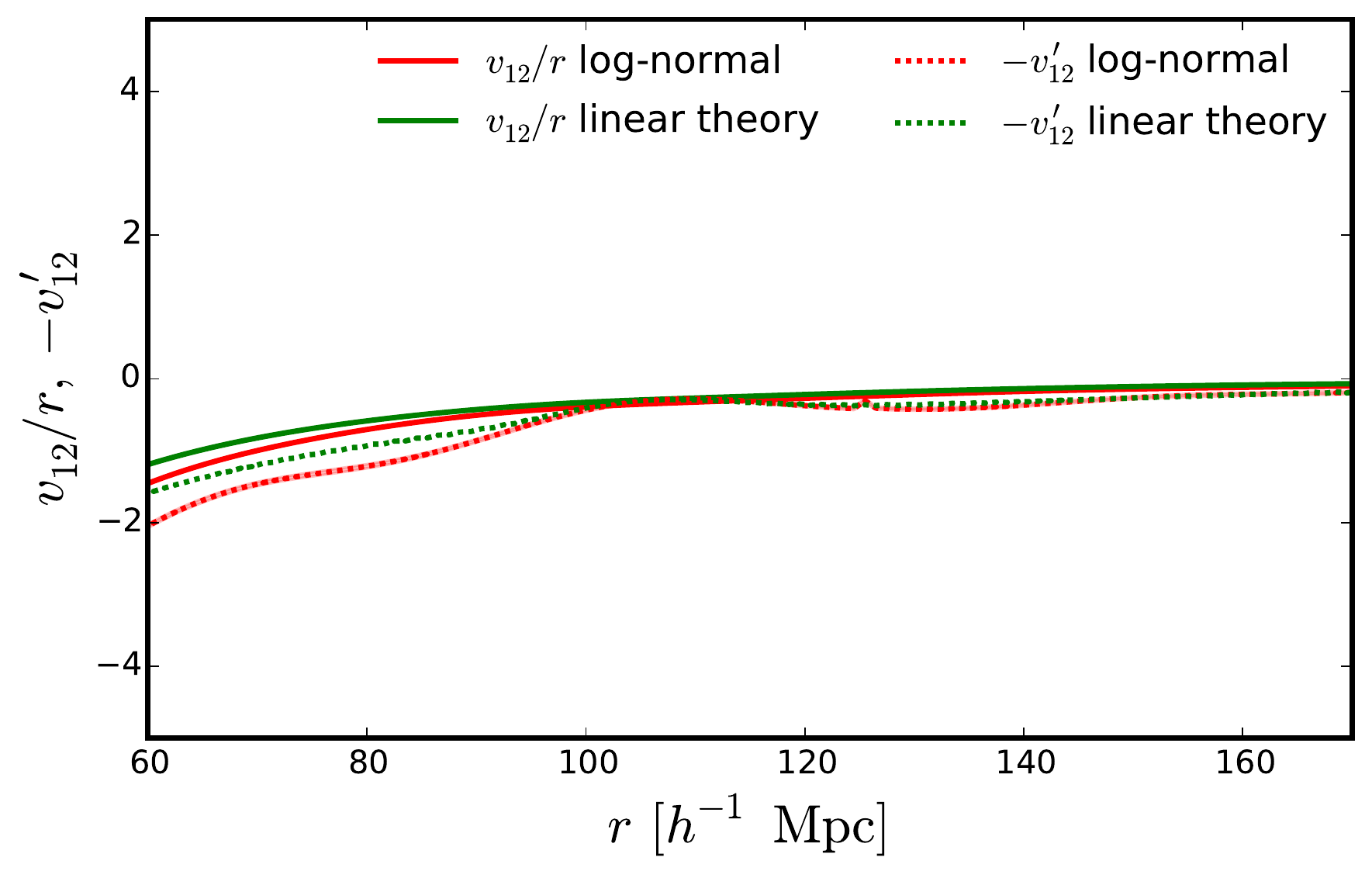}
	\caption{Mean radial pairwise velocity (solid) measured from our log-normal mocks (red) as well as that predicted by linear theory (green). Derivative w.r.t. separation $r$ (dotted). Red band denotes the error on the mean measured from 50 realisations. They agree at separations $\gtrsim 140\, h^{-1}$ Mpc. }
	\label{fig:vmean}
\end{figure}
%%%%%%%%%%%%%%%%%%%%%%%%%%%%%%%%%%%%%%%%%%%%%%%%%%%%%%%%%%%%%%%%%%%%%%%%

%%%%%%%%%%%%%%%%%%%%%%%%%%%%%%%%%%%%%%%%%%%%%%%%%%%%%%%%%%%%%%%%%%%%%%%%
\begin{figure}[t]
	\centering
	\includegraphics[width=1\textwidth, bb=0 0 541 333]{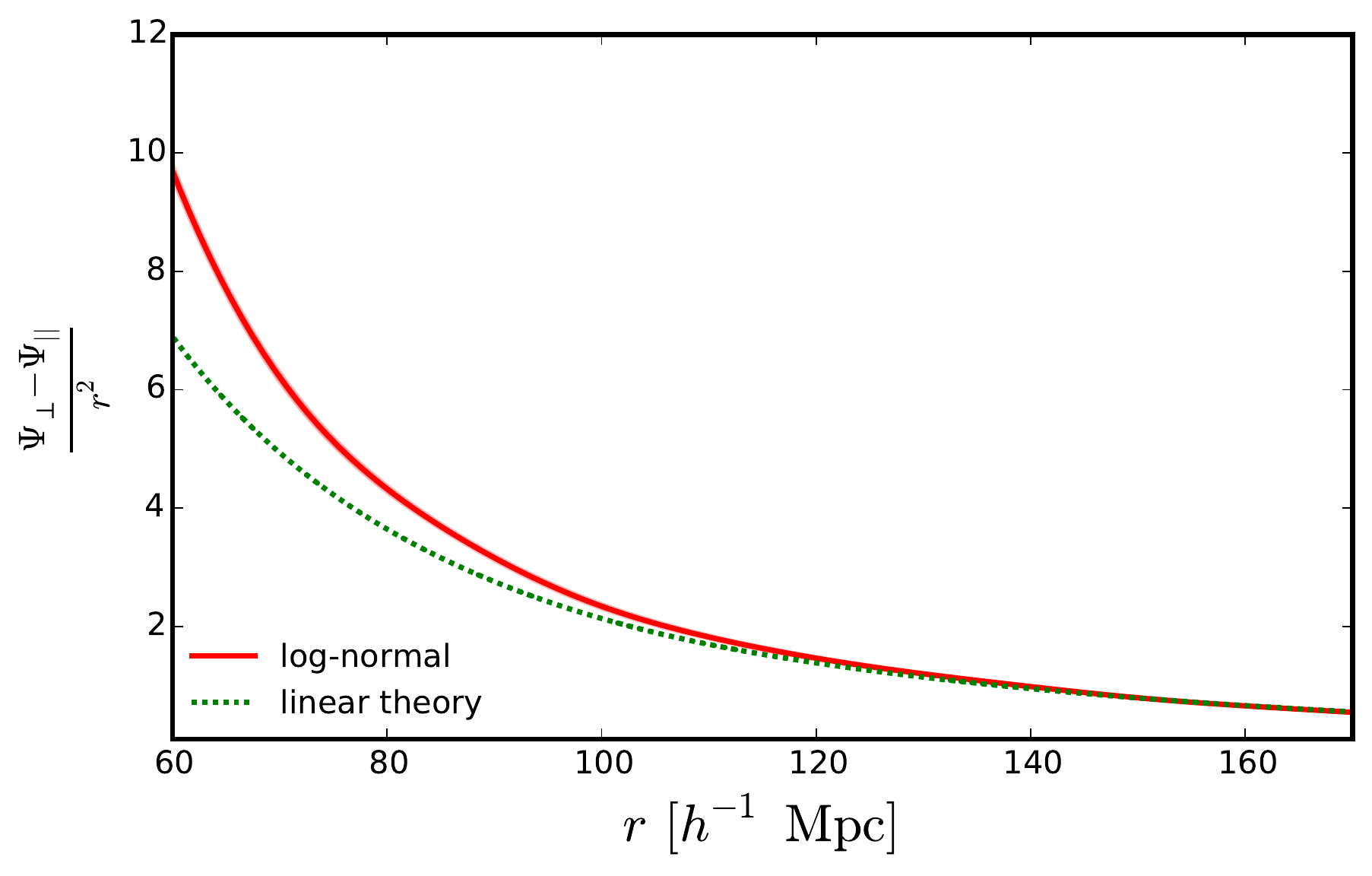}
	\caption{Same as \reffig{vmean} but for difference of radial ($\Psi_{\para}$) and tangential ($\Psi_{\perp}$) pairwise velocity variances. They agree at separations $\gtrsim 140\, h^{-1}$ Mpc.}
	\label{fig:psi_d}
\end{figure}
%%%%%%%%%%%%%%%%%%%%%%%%%%%%%%%%%%%%%%%%%%%%%%%%%%%%%%%%%%%%%%%%%%%%%%%%

%%%%%%%%%%%%%%%%%%%%%%%%%%%%%%%%%%%%%%%%%%%%%%%%%%%%%%%%%%%%%%%%%%%%%%%%
\begin{figure}[t]
	\centering
	\includegraphics[width=1\textwidth, bb=0 0 508 314]{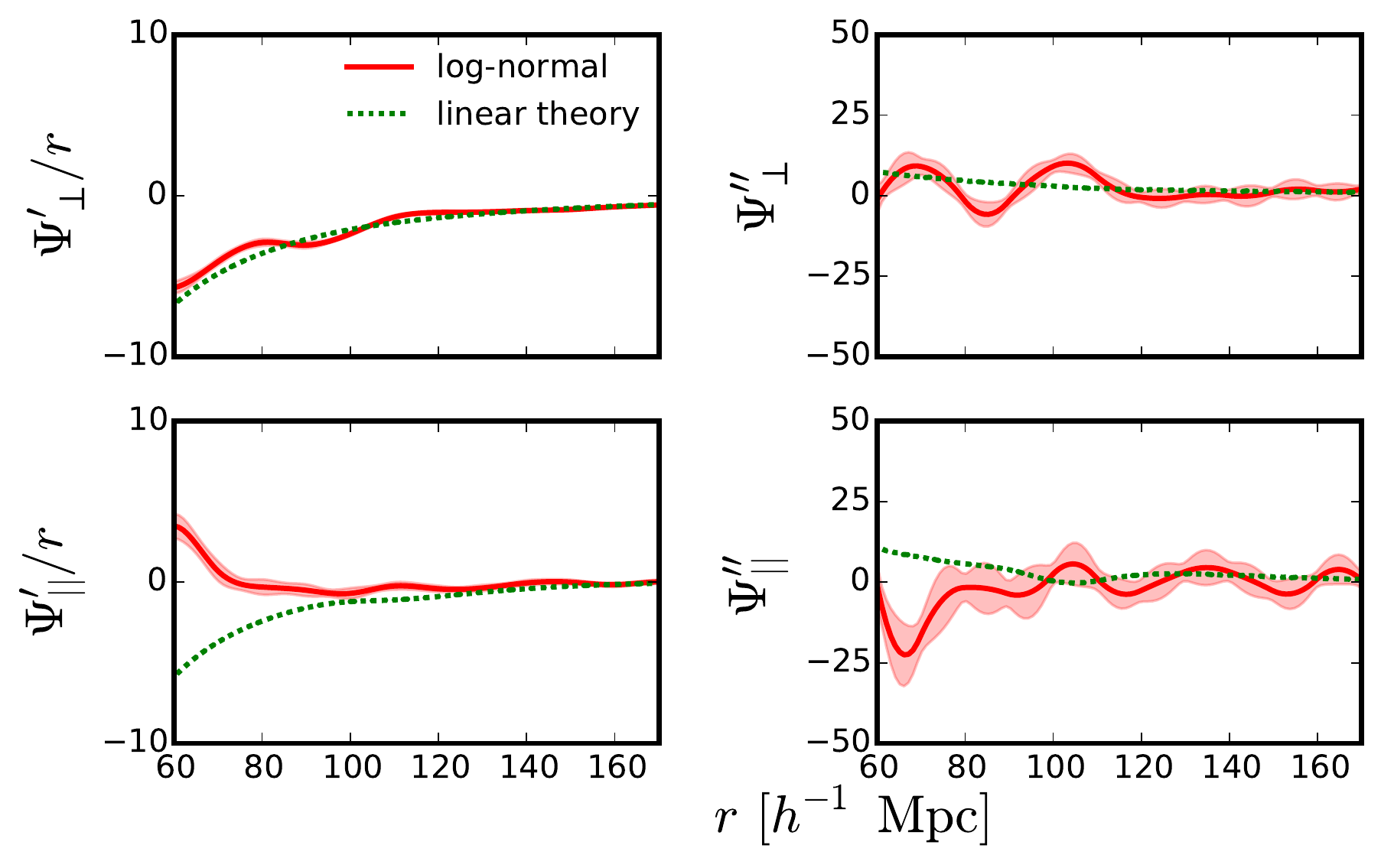}
	\caption{Same as \reffig{vmean} but for first and second derivatives of tangential and radial pairwise velocity variances. The second derivatives are noisy but they seem to agree with linear theory predictions at separations $\gtrsim 120\, h^{-1}$ Mpc. 
	}
	\label{fig:psi_derivs}
\end{figure}
%%%%%%%%%%%%%%%%%%%%%%%%%%%%%%%%%%%%%%%%%%%%%%%%%%%%%%%%%%%%%%%%%%%%%%%%

In \reffigs{vmean}{psi_derivs} we show the different terms that contribute to the Kaiser limit of the two-point correlation function, namely, $v_{12}/r$, $v'_{12}$, $(\Psi_{\perp}-\Psi_{\para})/r^2$, $\Psi'_{\para}/r$, $\Psi'_{\perp}/r$, $\Psi''_{\para}$, and $\Psi''_{\perp}$ for our log-normal mocks 
{(run to match the conditions from the N-body simulations, i.e. we choose $b=1.8$ which is the average linear bias for halos in the mass range $5 \times 10^{13} - 6.5 \times 10^{13}$ $h^{-1}\,M_{\odot}$ at $z=0$)} %, and the same cosmology : $\Omega_m=0.27$, $\Omega_{\Lambda}=0.73$, $n_s = 0.95$, $A = 2.2 \times 10^{-9}$)
and linear theory predictions. They agree at large separations, which ensures that the Kaiser limit is attained. This is because only the spatial derivatives of the first two moments of the pairwise velocity PDF contribute to the lowest-order redshift space correlation function in the large-scale limit  \cite{Fisher:1994ks,scoccimarro:2004,reid:2011}, as we confirm. So, even though the PDF from our log-normal mocks shows a non-zero excess kurtosis on large scales {(which is a consequence of the non-zero excess kurtosis of the log-normal density PDF, as only the one-point distribution is relevant for large separations)} and the dispersion is larger than the one from the N-body simulations (by a scale-independent constant), we find agreement with the Kaiser prediction on large scales. 

%%%%%%%%%%%%%%%%%%%%%%%%%%%%%%%%%%%%%%%%%%%%%%%%%%%%%%%%%%%%%%%%%%%%%%%%
\section{Summary and Conclusions}
\label{sec:summary}
%%%%%%%%%%%%%%%%%%%%%%%%%%%%%%%%%%%%%%%%%%%%%%%%%%%%%%%%%%%%%%%%%%%%%%%%
We have presented a new public code for generating log-normal realisations 
including velocity fields satisfying the linear continuity equation. 
The log-normal realisations provide not only a fast and easy way to generate 
mock galaxy catalogs but also an excellent test bed for studying
non-linear effects such as the window function and RSD. 

We have verified that the real-space two-point correlation functions measured 
from our log-normal mock galaxy catalogs are in excellent agreement with the 
input. We find that the cross-correlation coefficients between
the matter and galaxy density fields are not unity
\cite{Xavier:2016elr}. Non-linear (exponential, to be specific)
transformation of perfectly correlated Gaussian fields induces a
deviation of the cross-correlation coefficient from unity. We
analytically compute the cross-correlation coefficient that matches the
measurement to a sub-percent level.

We have also shown measurements from our log-normal mock catalogs in redshift 
space. The redshift-space power spectrum is commonly modelled as a combination 
of a ``squashing'' term (in the Kaiser limit, arising from coherent 
large-scale flows) and a damping term (FoG from random virial 
motion on small scales). Using our log-normal mock catalogs, we have 
investigated the redshift-space power spectrum and found a good agreement 
with the squashing term on large scales ($k \apprle 0.1\, h^{-1}$ Mpc) as 
expected. On small scales, we find a damping which is qualitatively 
similar to the FoG; the damping we observe, however, does 
not come from random motion, as we do not include any random virial motion 
in our mock generator. Rather, the damping comes from non-linearity in the
Jacobian of the real-to-redshift space mapping, and from the coherent peculiar
velocity field. Attributing all of the damping to the FoG, as commonly
done in the literature, is thus misleading. The configuration space
two-point correlation function calculated with the linear Jacobian
approximation cannot reproduce the measurement at all separations; thus,
the correlation function is sensitive to non-linearity in the Jacobian
even at large separations.

The streaming model can take into account the full non-linearity of
the real-to-redshift space mapping. In this model a fundamental entity
in predicting the redshift-space two-point statistics is the pairwise
line-of-sight velocity PDF, which is notoriously hard to predict owing
to its {\it pairwise} nature. We find that the problem persists even
with our, rather simpler, setting: the PDF of the density field is
exactly known to be log-normal, and the velocity is linearly related to
the density field. We nevertheless have made some progress in a couple of areas in modelling the pairwise line-of-sight velocity PDF. First, we show that the pairwise line-of-sight velocity PDF from our log-normal mock catalogs qualitatively captures features of the PDF from full N-body simulations, such as a negative 
skewness for small separations and the shift of the PDF towards more negative 
velocities for higher mass halos. We find these features
even when the same coherent velocity field is assigned 
to galaxy fields with different biases, i.e., galaxies with different
biases move with the same velocities, but pair-weighting makes the
pairwise velocity PDF depend on the galaxy bias.
Second, for the log-normal setting, one can in principle predict the moments 
of the pairwise line-of-sight velocity PDF, as we explicitly demonstrate
for the mean pairwise line-of-sight velocity. We have compared the predicted 
mean velocity to the one measured from the catalogs, and find an excellent 
match between the two. Likewise, although very demanding, we envisage that 
the analytical calculation can also be done for the higher order moments.

Our log-normal generator has been used extensively to help design the
on-going and planned galaxy redshift surveys such as HETDEX
(Hobby-Eberly Telescope Dark Energy Experiment) \cite{Hill:2008mv}, PFS
(Prime Focus Spectrograph) \cite{Tamura:2016wsg}, and WFIRST-AFTA (Wide Field
Infrared Survey Telescope Astrophysics-Focused Telescope Assets)
\cite{Spergel:2015sza}. It should also be equally
useful for DESI (Dark Energy Spectroscopic Instrument)
\cite{Levi:2013gra}, LSST (Large Synoptic Survey Telescope)
\cite{abell2009lsst}, and Euclid \cite{Laureijs:2011gra}. In near future
we shall make available a code for computing weak gravitational lensing
fields from the log-normal density field (``log-normal\_lens''; Makiya et al., in
preparation). This code allows us to study the cross-correlation power
spectrum between galaxy positions and weak lensing fields, which is one
of the products of PFS and Euclid as well as LSST with spectroscopic
follow-ups. 

\acknowledgments
We would like to thank I. Jee, I. Kayo, and F. Schmidt for useful
discussions, and G.~E. Addison, C.~L. Bennett, and J.~L. Weiland for
comments on the draft. This work was supported in part by MEXT KAKENHI
Grant Number 15H05896. CC is supported by grant NSF
PHY-1620628. D.J. was supported by National Science Foundation grant
AST-1517363. We also acknowledge NASA grant NNX15AJ57G.

%%%%%%%%%%%%%%%%%%%%%%%%%%%%%%%%%%%%%%%%%%%%%%%%%%%%%%%%%%%%%%%%%%%%%%%%%%%%
\appendix
%%%%%%%%%%%%%%%%%%%%%%%%%%%%%%%%%%%%%%%%%%%%%%%%%%%%%%%%%%%%%%%%%%%%%%%%%%%%

%%%%%%%%%%%%%%%%%%%%%%%%%%%%%%%%%%%%%%%%%%%%%%%%%%%%%%%%%%%%%%%%%%%%%%%%%%%%
\section{Derivation of the streaming model in configuration space}
\label{app:streaming}
%%%%%%%%%%%%%%%%%%%%%%%%%%%%%%%%%%%%%%%%%%%%%%%%%%%%%%%%%%%%%%%%%%%%%%%%%%%%
In this appendix, we re-derive the streaming model 
\cite{peebles:book,Davis:1982gc,Fisher:1994ks,scoccimarro:2004,uhlemann:2015},
which equates the redshift-space two-point correlation function 
$\xi_{gg}^s$ to the real-space two-point correlation function $\xi_{gg}$
re-mapped by the pairwise line-of-sight velocity PDF
${\cal P}$ as 
\be
1+\xi_{gg}^s(\vecs) = 
\int  d  r_\parallel
{\cal P} \left( s_\parallel- r_\parallel; \vr \right)
\left[1+\xi_{gg}(\vr)\right]\,.
\label{eq:streaming_app}
\ee
Here, $\vecs$ and $\vr$ are, respectively, the separations in 
redshift space and real space.
To the best of our knowledge, the streaming model in the form of 
\refeq{streaming_app} has first appeared in \cite{peebles:book}, and 
the later studies \cite{Davis:1982gc,Fisher:1994ks,scoccimarro:2004,uhlemann:2015}
have improved the modelling and interpretation of the pairwise line-of-sight 
velocity PDF. For example, Ref.~\cite{Fisher:1994ks} incorporates the 
scale-dependence of the velocity dispersion to reproduce the Kaiser 
\cite{kaiser:1987} prediction; Ref.~\cite{scoccimarro:2004} finds the 
expression for the pairwise line-of-sight velocity PDF and its moment 
generating function with an assumption that the velocity field $\vv(\vr)$ is 
a single-valued function of positions (we shall call this {\it single-stream} 
case). More recently, Ref.~\cite{uhlemann:2015} generalizes the results 
to the {\it multi-stream} case where there are multiple velocity components 
(streams) at a single position; this is, for example, the case for the shell 
crossing in the spherical collapse.
Here, we shall closely follow the result of Ref.~\cite{uhlemann:2015} so that 
derivation we present here works for the multi-streaming case.

Starting from the galaxy number conservation between the real and redshift 
space (\refeq{conservation}), and using the general phase space function 
$f(\vx,\vv)$,
we may write the redshift-space density contrast as \cite{Seljak:2011tx}:
\ba
1+\delta_g^s(\vecs)
=
\int d^3x\int d^3v~f(\vx,\vv)
\delta^D\left(\vx +\frac{v_\parallel}{\cH}\hat\ell - \vecs\right),
\label{eq:constrastrsd}
\ea
where $\delta^D$ is the Dirac-delta operator.
The redshift-space two-point correlation function is then given as
\ba
&\,\left<\left[1+\delta_g^s(\vecs_1)\right]\left[1+\delta_g^s(\vecs_2)\right]\right>
= \nonumber \int d^3\vx_1 \int d^3\vx_2 \int d^3\vv_1 \int d^3\vv_2
\\
&\,\left<
\delta^D\left(\vx_1-\vecs_1+\frac{v_{1,\para}}{\cH}\hat{\ell}\right)
\delta^D\left(\vx_2-\vecs_2+\frac{v_{2,\para}}{\cH}\hat{\ell}\right)
f(\vx_1,\vv_1)
f(\vx_2,\vv_2)
\right>.
\label{eq:rsd_pair_conserv}
\ea
As is apparent from \refeq{r_to_s}, the RSD only applies to the line-of-sight 
quantities, so it is helpful to explicitly indicate the 
line-of-sight quantities with the subscript $\para$ and 
perpendicular quantities with the subscript $\perp$.
Then, \refeq{rsd_pair_conserv} becomes 
\ba
\nonumber & \left<
\left[1+\delta_g^s\left(s_{1,\para},s_{1,\perp}\right)\right]\left[1+\delta_g^s\left(s_{2,\para},s_{2,\perp}\right)\right]
\right> = \int dx_{1,\para} \int dx_{2,\para} \int d^3\vv_1 \int d^3\vv_2
\\ 
&\left<
\delta^D\left(x_{1,\para}-s_{1,\para}+\frac{v_{1,\para}}{\cH}\right)
\delta^D\left(x_{2,\para}-s_{2,\para}+\frac{v_{2,\para}}{\cH}\right)
f(x_{1,\para},s_{1,\perp},\vv_1)
f(x_{2,\para},s_{2,\perp},\vv_2)
\right>\,.
\label{eq:rsd_pair_conserv_perp_para}
\ea
Here, we keep the Dirac-delta operators inside the ensemble average as 
they contain the peculiar velocity field.
We then use the definition of the Dirac-delta
$\delta^D(x) = (2\pi)^{-1}\int_{-\infty}^\infty d\gamma e^{-i\gamma x}$,
to transform \refeq{rsd_pair_conserv_perp_para} as
\ba
%&\,
&\left<
\left[1+\delta_g^s\left(s_{1,\para},s_{1,\perp}\right)\right]\left[1+\delta_g^s\left(s_{2,\para},s_{2,\perp}\right)\right]
\right>
%\vs
=\nonumber
\int dx_{1,\para}
\int dx_{2,\para}
\int \frac{d\gamma_1}{2\pi}
\int \frac{d\gamma_2}{2\pi}
e^{-i\gamma_1\left(x_{1,\para}-s_{1,\para}\right)}
\\
&
e^{-i\gamma_2\left(x_{2,\para}-s_{2,\para}\right)}
\times %\quad\qquad\quad\times
\int d^3\vv_1
\int d^3\vv_2
\left<
e^{-i\gamma_1\frac{v_{1,\para}}{\cH}}
e^{-i\gamma_2\frac{v_{2,\para}}{\cH}}
f(x_{1,\para},s_{1,\perp},\vv_1)
f(x_{2,\para},s_{2,\perp},\vv_2)
\right>.
\ea
Because of statistical homogeneity of the Universe, the ensemble average 
must depend only on the separation. We make it explicit by introducing new 
variables $R_\para = (x_{1,\para}+x_{2,\para})/2$ and 
$r_\para = x_{1,\para}-x_{2,\para}$,
with which the right-hand side of the equation above becomes
\ba
&\,
\left<
\left[1+\delta_g^s\left(s_{1,\para},s_{1,\perp}\right)\right]\left[1+\delta_g^s\left(s_{2,\para},s_{2,\perp}\right)\right]
\right>
\vs
=&\,
\int d r_\para
\int \frac{d\gamma_1}{2\pi}
\int d\gamma_2
\left[
\int \frac{dR_\para}{2\pi}
e^{-i R_\para\left(\gamma_1 + \gamma_2\right)}
\right]
e^{- \frac{i}{2} r_\para\left(\gamma_1 - \gamma_2\right)}
e^{i\left(\gamma_1 s_{1,\para} + \gamma_2 s_{2, \para}\right)}
\vs
&\times
\int d^3\vv_1
\int d^3\vv_2
\left<
e^{-i\gamma_1\frac{v_{1,\para}}{\cH}}
e^{-i\gamma_2\frac{v_{2,\para}}{\cH}}
f(x_{1,\para},s_{1,\perp},\vv_1)
f(x_{2,\para},s_{2,\perp},\vv_2)
\right>
\vs
=&\,
\int d r_\para
\int \frac{d\gamma_1}{2\pi}
\int d\gamma_2
\delta^D(\gamma_1+\gamma_2)
e^{- \frac{i}{2}r_\para\left(\gamma_1 - \gamma_2\right)}
e^{i\left(\gamma_1 s_{1,\para} + \gamma_2 s_{2, \para}\right)}
\vs
&\times
\int d^3\vv_1
\int d^3\vv_2
\left<
e^{-i\gamma_1\frac{v_{1,\para}}{\cH}}
e^{-i\gamma_2\frac{v_{2,\para}}{\cH}}
f(x_{1,\para},s_{1,\perp},\vv_1)
f(x_{2,\para},s_{2,\perp},\vv_2)
\right>\,.
\ea
Finally, integrating the Dirac-delta yields
\ba
1+\xi_{gg}^s(s_\para, s_\perp)
=&\,
\int d r_\para
\int \frac{d\gamma_1}{2\pi}
e^{- i\gamma_1 (r_\para - s_\para)}
\int d^3\vv_1
\int d^3\vv_2
\left<
e^{-i\gamma_1\frac{\Delta v_\para}{\cH}}
f(x_{1,\para},s_{1,\perp},\vv_1)
f(x_{2,\para},s_{2,\perp},\vv_2)
\right>\,,
\ea
with $\Delta v_\para = v_{1,\para}-v_{2,\para} $.
Again, note that the ensemble average must depend only on the separation.
Following Ref.~\cite{scoccimarro:2004}, we define the pairwise line-of-sight
velocity PDF as
\be
{\cal P}(r_\parallel-s_\parallel,\vr)
=
\int \frac{d\gamma}{2\pi} e^{-i\gamma (r_\parallel- s_\parallel)}
{\cal M}(-i\gamma , \vr),
\ee
where $\vr=\vx_1-\vx_2$ and ${\cal M}(\lambda,\vr)$ is the generating function associated with 
the pairwise line-of-sight velocity PDF:
\be
\left[1+\xi_{gg}( \vr)\right]{\cal M}(\lambda, \vr)
\equiv
\int d^3\vv_1
\int d^3\vv_2
\left<
e^{\lambda\frac{\Delta v_\para}{\cH}}
f(\vx_1,\vv_1)
f(\vx_2,\vv_2)
\right>\,.
\label{eq:MM}
\ee
This leads to the streaming model:
\be
1+\xi_{gg}^s(s_\para,s_\perp)
=\,
\int d r_\para
{\cal P}(r_\parallel - s_\parallel;\vr)
\left[
1+ \xi_{gg}(\vr)
\right],
\ee
where $\vr_{\perp} = \vecs_{\perp}$, and the real space two-point correlation function is given as
\be
1+ \xi_{gg}(\vr)
\equiv 
\int d^3\vv_1
\int d^3\vv_2
\left<
f(\vx_1,\vv_1)
f(\vx_2,\vv_2)
\right>\,.
\ee
Along the course of the derivation, we have only used the homogeneity of the
Universe. We, therefore, conclude that the streaming model is an {\it exact}
expression for the redshift-space two-point correlation function, following 
from the number conservation and the statistical homogeneity. 

Note that for the single streaming case, where the distribution function 
may be written as 
$f(\vx,\vv) = \left[1+\delta_g(\vx)\right]\delta^D(\vv-\bar{\vv}(\vx))$
with the bulk velocity $\bar\vv(\vx)$ uniquely defined at the position $\vx$,
\refeq{MM} reduces to the the result of \cite{scoccimarro:2004}:
\be
\left[1+\xi_{gg}( \vr)\right]{\cal M}(\lambda, \vr)
\stackrel{\rm single~stream}{=}
\left<
e^{\lambda\Delta \bar{v}_\parallel/\cH}
\left[
1+\delta_g(\vx_1)
\right]
\left[
1+\delta_g(\vx_2)
\right]
\right>.
\ee
The general formula, \refeq{MM}, must be used whenever multiple velocities
are assigned to single spatial elements. That happens, for example, when 
coarse-graining the galaxy density field.

%%%%%%%%%%%%%%%%%%%%%%%%%%%%%%%%%%%%%%%%%%%%%%%%%%%%%%%%%%%%%%%%%%%%%%%%%%%%
\section{Binning effect of the power spectrum measurement}
\label{app:pk_bin}
%%%%%%%%%%%%%%%%%%%%%%%%%%%%%%%%%%%%%%%%%%%%%%%%%%%%%%%%%%%%%%%%%%%%%%%%%%%%
For a density field in a cubic volume of $V=L^3$, we estimate the power 
spectrum at $k=nk_F$ ($n$ is an integer and $k_F\equiv 2\pi/L$ is the
fundamental  wavenumber) by taking the average over the amplitudes of
Fourier modes around $k$ \cite{Feldman:1993,djeong_thesis:2010}:
\be
 P(k_F n) = \frac{V}{N^6}\left( \frac{1}{N_k}\sum_{\left| n_k-n\right| \leq 1/2} \left|
 \delta_{\rm FFTW}(\textbf{n}_k)\right| ^2\right) \,,
\label{eq:pk_est}
\ee
where $\delta_{\rm FFTW}$ is the density field in Fourier space, 
$N$ is the number of one-dimensional grid so that $H^3 = V/N^3$ becomes the 
volume of one grid Fourier cell, 
and $N_k$ is the number of discrete Fourier modes falling into the bin.
Because of the binning, the estimated power spectrum at $k$ in 
\refeq{pk_est} may differ from the true power spectrum $P(nk_F)$;
we call it a binning effect.
This effect is particularly important on large scales, where the number of 
Fourier modes is small. To make accurate comparison between the measurement 
and prediction, we need to take this effect into account. In the following, 
we explore three methods to account for the binning effect.

\begin{enumerate}
\item{
Compute the prediction by volume-averaging the input power spectrum $P_{\rm inp}(k)$, i.e.,
\be
 P_{\rm smooth}(k) = \frac{\int_{k_{\rm min}}^{k_{\rm max}} dk k^2~P_{\rm inp}(k)}{\left( k^3_{\rm max}-k^3_{\rm min}\right)/3 } \,,
\ee
where $k_{\rm max}$ and $k_{\rm min}$ denote the boundaries of the particular $k$ bin.
We shall refer to this as ``smoothed''. 
}
\item{
Volume-average the wavenumber to compute an \textit{effective} wavenumber for each $k$ bin
\be
 k_{\rm eff} = \frac{\int_{k_{\rm min}}^{k_{\rm max}} dk k^2~k}{\left( k^3_{\rm max}-k^3_{\rm min}\right)/3 }
 = \frac{3}{4}\frac{\left( k^4_{\rm max}-k^4_{\rm min}\right)}{\left( k^3_{\rm max}-k^3_{\rm min}\right)} \,,
\ee
and interpolate the input power spectrum at this effective wavenumber $P(k_{\rm eff})$.
We shall refer to this as ``$k$-smoothed''. 
}
\item{
Interpolate the input power spectrum
on each $\vk$ grid, and then bin this interpolated power spectrum. Namely,
\ba
 P_{\rm discrete}(k_F \textbf{n}_1)
 = \frac{V}{N^6}\left( \frac{1}{N_k}\sum_{\left| n_k-n_1\right| \leq 1/2} P_{\rm inp}(\textbf{n}_kk_F)\right)\,,
\label{eq:pk_disc}
\ea
and we shall refer to this as ``discrete''.
}
\end{enumerate}

\begin{figure}[t]
\centering
\includegraphics[width=0.9\textwidth, bb = 0 0 522 321]{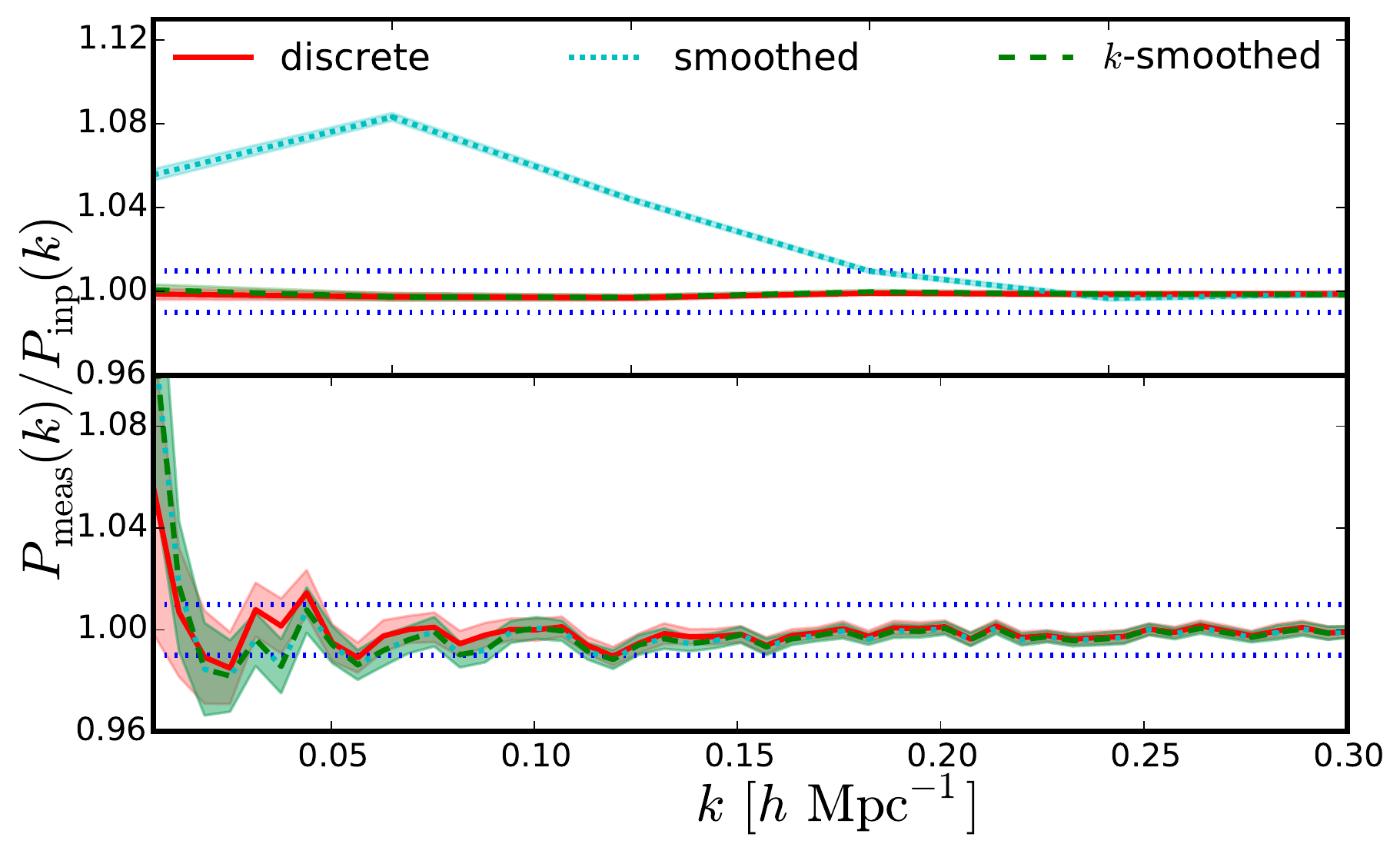}
\caption{Ratio of the measured power spectrum to the input for three different
methods of accounting for the binning effect. (Top) Bin size of $0.05\ihMpc$. (Bottom)
Bin size of $0.006\ihMpc$, which is the fundamental frequency in our mock
catalog. The
three methods are shown in the red solid (discrete), cyan dotted
 (smoothed), and green dashed ($k$-smoothed) lines. The band denotes the error on the mean measured from 50
realisations. The cyan and green bands overlap in the bottom panel.}
\label{fig:pk_bin}
\end{figure}

\refFig{pk_bin} shows the ratio of the measured power spectrum to the
input power spectrum computed using the above three methods. The top and bottom
panels show binning sizes of 0.05 and $=0.006\ihMpc$ (which is the fundamental
frequency), respectively. For the large $k$ bin, the smoothed method is
inaccurate but the $k$-smoothed and discrete methods agree well with
the measurement; for the small $k$ bin, all methods perform similarly,
with the discrete method performing slightly better at
$k\lesssim0.02\ihMpc$. Thus, in this paper we shall use the discrete
method for computing the prediction.

%%%%%%%%%%%%%%%%%%%%%%%%%%%%%%%%%%%

%%%%%%%%%%%%%%%%%%%%%%%%%%%%%%%%%%%%%%%%%%%%%%%%%%%%%%%%%%%%%%%%%%%%%%%%%%%%
\section{Mean pairwise line-of-sight velocity in log-normal mock catalog}
\label{app:mean_pairwise_v}
%%%%%%%%%%%%%%%%%%%%%%%%%%%%%%%%%%%%%%%%%%%%%%%%%%%%%%%%%%%%%%%%%%%%%%%%%%%%
From \refeq{mgf_0}, we find that the mean of the pairwise line-of-sight velocity
is given by
\be
 \left[ 1+\xi_{gg}(r)\right] \left\langle \Delta v_z\right\rangle
 =\left\langle v_{1z}\delta_{g2}\right\rangle
 -\left\langle v_{2z}\delta_{g1}\right\rangle
 +\left\langle v_{1z}\delta_{g1}\delta_{g2}\right\rangle
 -\left\langle v_{2z}\delta_{g2}\delta_{g1}\right\rangle \,,
\label{eq:mgf_0_expanded}
\ee
where we use $\left\langle v_{1z}\right\rangle= \left\langle v_{2z}\right\rangle$
and $\left\langle v_{1z}\delta_{g1}\right\rangle = \left\langle
v_{2z}\delta_{g2}\right\rangle$ from homogeneity and isotropy. Thus, to compute
the mean of the pairwise line-of-sight velocity, we need the contributions
from both two- and three-point functions.

\begin{figure}[t]
\centering
\includegraphics[width=0.4\textwidth, bb=0 0 533 333]{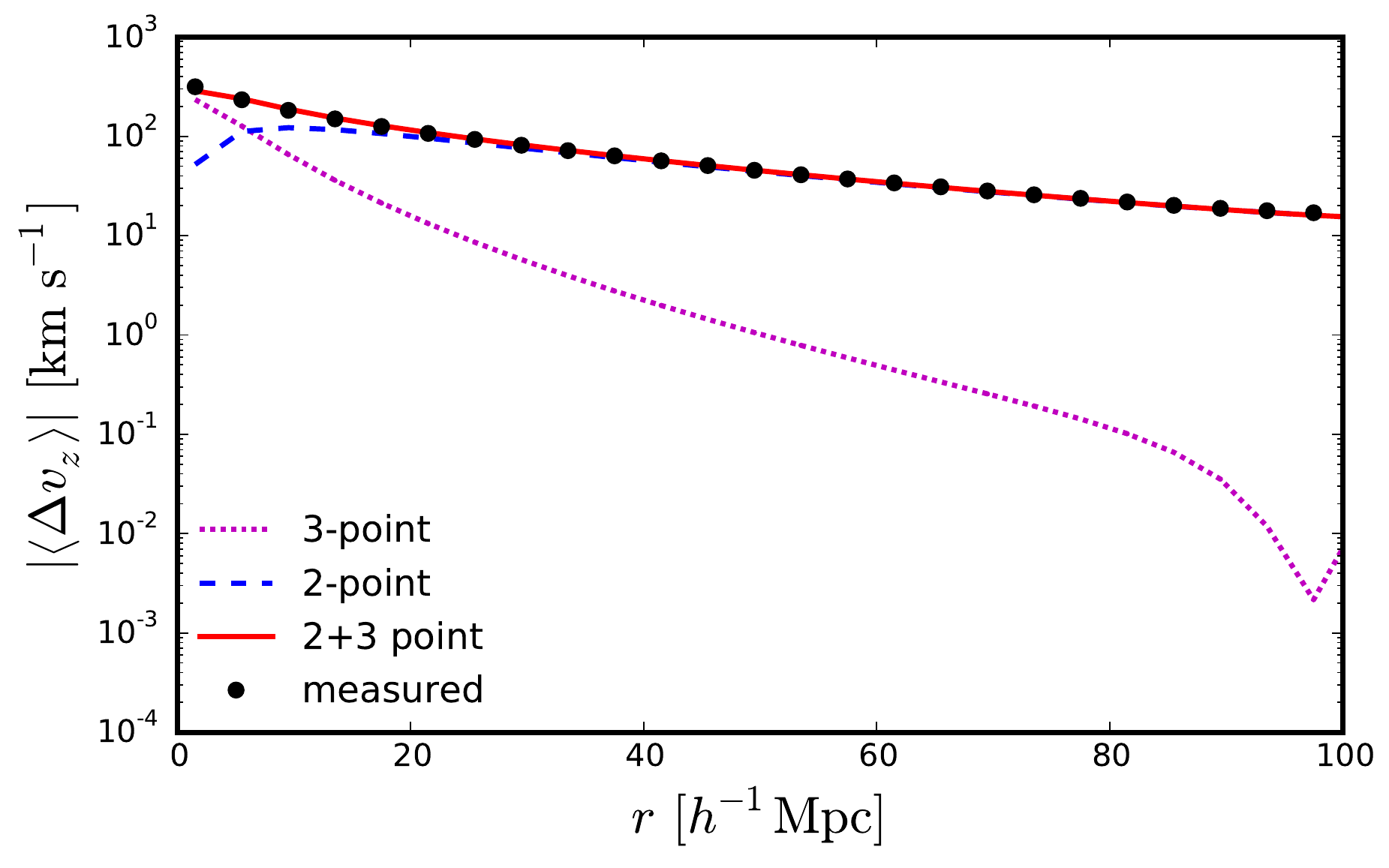}
\includegraphics[width=0.4\textwidth, bb=0 0 527 331]{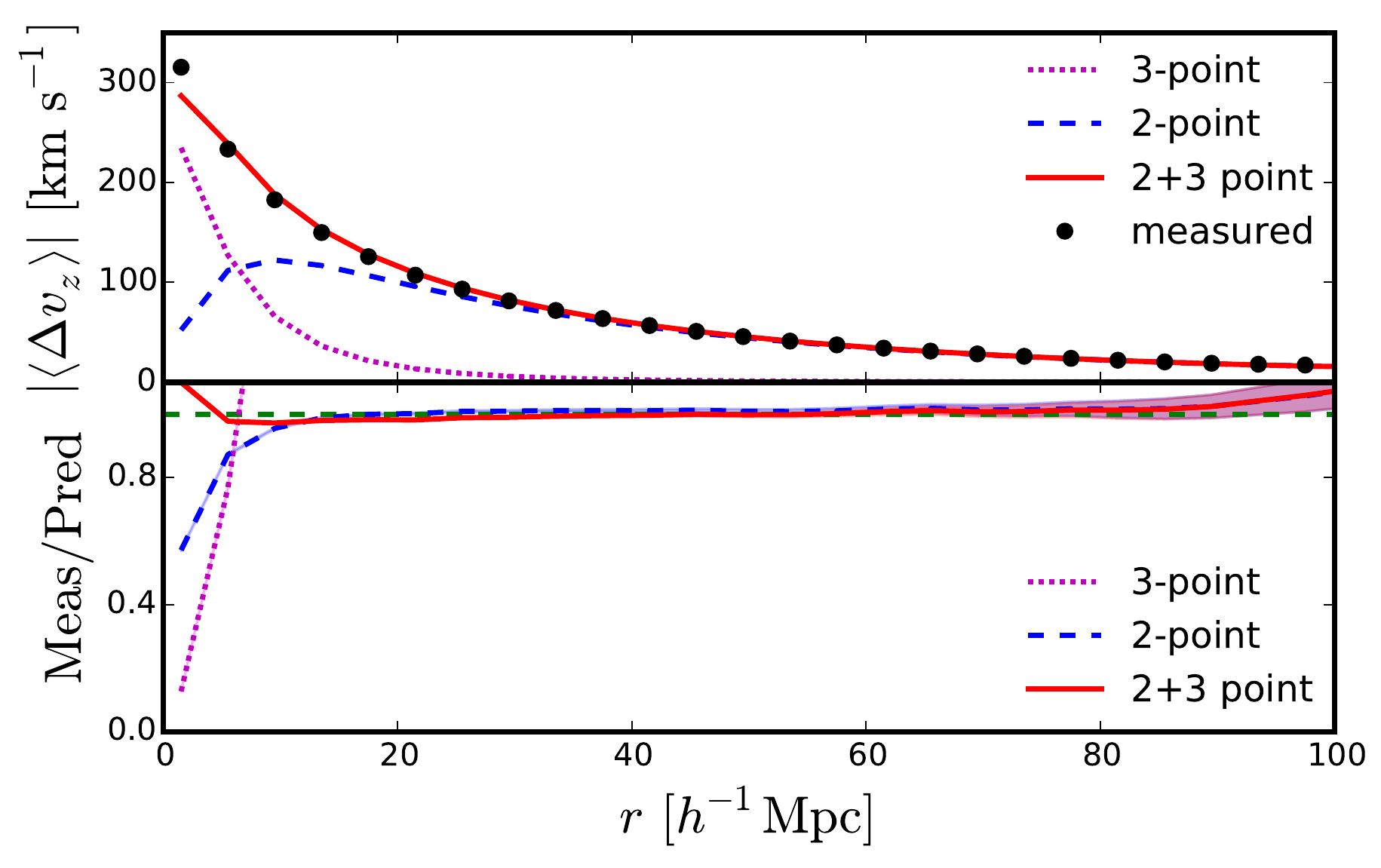}
\caption{{Calculation of the mean pairwise line-of-sight ($\mu = 0.995$) velocity
in our log-normal mock catalogs, in log scale (left) and linear scale (right, top), along 
with the residuals (right, bottom). The magenta dotted, blue dashed, and red solid lines
show the contributions from three-point function alone, two-point function alone, 
and two- and three-point functions, respectively. The black points show the measurement
from the log-normal mock catalogs and the green dashed line in the bottom right panel shows 1.}}
\label{fig:mean_vel}
\end{figure}

The two-point function contribution is given by
\be
 \left\langle v_{iz}\delta_{gj}\right\rangle
 =i\mathcal{H}f \int \frac{d^3k}{(2\pi)^3}\frac{k_{z}}{k^2}P_{gm}(k)e^{i\vk\cdot\vr_{ij}}
 =i\mathcal{H}f \mu \int \frac{dk}{2\pi^2}kP_{gm}(k)j_1(kr_{ij}) \,,
\label{eq:v_del}
\ee
where $\vr_{ij}\equiv \vx_i-\vx_j$, $\mu\equiv \hat{z}\cdot\hat{r}_{ij}$ and $j_1(x)$ is the spherical Bessel function
of the first order. Note that this product is anti-symmetric under the exchange
of $i$ and $j$, hence
$\left\langle v_{1z}\delta_{g2}\right\rangle - \left\langle v_{2z}\delta_{g1}\right\rangle = 2 \left\langle v_{1z}\delta_{g2}\right\rangle$.
Using \refeqs{sph_ft}{pGgm}, \refeq{v_del} can be evaluated numerically as a
function of $r$ and $\mu$. The blue dashed line in \reffig{mean_vel} shows its
contribution to the mean pairwise line-of-sight velocity for $\mu = 0.995$. We
find that as the separation approaches to zero, this contribution drops to zero
since $j_1(x) \to 0$ for $x \to 0$. 
The contribution also decreases with increasing separation, which is 
a generic feature following the trend of the density two-point correlation 
function.

%%%%%%%%%%%%%%%%%%%%%%%%%%%%%%%%%%%%%%%%%%%%%%%%%%%%%%%%%%%%%%%%%%%%%%%%%%%%%

%%%%%%%%%%%%%%%%%%%%%%%%%%%%%%%%%%%%%%%%%%%%%%%%%%%%%%%%%%%%%%%%%%%%%%%%%%%%%

For the three-point function contribution, we have
\be
 \left\langle v_{iz}\delta_{gi}\delta_{gj}\right\rangle
 = i\mathcal{H}f \int d^3x_p\frac{d^3k}{(2\pi)^3}\frac{k_{z}}{k^2}
 e^{i\vk\cdot\vr_{ij}}\left\langle \delta_m(\vx_p)\delta_g(\vx_i)\delta_g(\vx_j)\right\rangle \,,
\label{eq:v_del_del}
\ee
which is an integral over the three-point function of the (matter and galaxy) density
fields. As the velocity field is linearly related to the density field in
Fourier space and the three-point function is only calculated easily in configuration space, we
need to introduce another vector variable $\vx_p$ to evaluate this contribution.\footnote{It
follows that for each power of velocity in the moment, we need to introduce one vector
variable and integrate over this variable, which makes this calculation impractical
for higher moments.} If the density fields are Gaussian, then this term vanishes and
we do not get any contribution. However, for log-normal fields this term is non-zero
and is given by \cite{coles:1991}
\begin{align}
 \left\langle \delta_m(\vx_p)\delta_g(\vx_i)\delta_g(\vx_j)\right\rangle
 =\:& \xi_{gm}(r_{pi})\xi_{gm}(r_{pj})\xi_{gg}(r_{ij}) + \vs
 \:& \xi_{gm}(r_{pi})\xi_{gg}(r_{ij})+\xi_{gm}(r_{pj})\xi_{gg}(r_{ij})
 +\xi_{gm}(r_{pi})\xi_{gm}(r_{pj}) \,,
\label{eq:3pf}
\end{align}
which can be evaluated for our mock catalogs. The magenta dotted line in \reffig{mean_vel}
shows this contribution. We find that the contribution decreases with increasing
separation, with an upturn at around the BAO scale. The two-point
function contribution dominates on most scales except on very small scales.

Combining the contributions from both two- and three-point functions, we find
good agreement between the analytic prediction and the measurement in the
log-normal mock catalogs, as demonstrated in \reffig{pdf_ln}.

%%%%%%%%%%%%%%%%%%%%%%%%%%%%%%%%%%%%%%%%%%%%%%%%%%%%%%%%%%%%%%%%%%%%%%%%%%%%%

%%%%%%%%%%%%%%%%%%%%%%%%%%%%%%%%%%%%%%%%%%%%%%%%%%%%%%%%%%%%%%%%%%%%%%%%%%%%
\section{Code Documentation}
\label{app:docu}
%%%%%%%%%%%%%%%%%%%%%%%%%%%%%%%%%%%%%%%%%%%%%%%%%%%%%%%%%%%%%%%%%%%%%%%%%%%%

%%%%%%%%%%%%%%%%%%%%%%%%%%%%%%%%%%%%%%%%%%%%%%%%%%%%%%%%%%%%%%%%%%%%%%%%%%%%
\subsection{Overview}
%%%%%%%%%%%%%%%%%%%%%%%%%%%%%%%%%%%%%%%%%%%%%%%%%%%%%%%%%%%%%%%%%%%%%%%%%%%%
Our mock generator code generates mock galaxy catalogs in redshift
space, assuming that the galaxy and matter density fluctuations of the
Universe follow a log-normal distribution. Since the code is
computationally inexpensive, one can easily generate plentiful
realisations with a large volume and a vast number of galaxies. This
property is convenient for studying systematics such as the survey
window function as well as to evaluate a covariance matrix of the galaxy
power spectrum for the current and planned large-scale structure
surveys.

%%%%%%%%%%%%%%%%%%%%%%%%%%%%%%%%%%%%%%%%%%%%%%%%%%%%%%%%%%%%%%%%%%%%%%%%%%%%
\subsection{Details}
\subsubsection{Input file}
%%%%%%%%%%%%%%%%%%%%%%%%%%%%%%%%%%%%%%%%%%%%%%%%%%%%%%%%%%%%%%%%%%%%%%%%%%%%
To run the code, one has to first prepare the configuration file
which should contain all the specifications of the model.
In this file one needs to specify the input cosmological parameters,
volume, output redshift, galaxy bias, number of galaxies, etc.
One also needs to specify some key parameters for the execution mode
of the code (e.g., the number of realisations, the number of parallel threads
to use, the choice of the power spectrum estimator, etc.).
The code automatically calculates the input matter power spectrum
with the cosmological parameters specified in the configuration file,
using the transfer function provided by Eisenstein \& Hu \cite{eisenstein/etal:1997,Eisenstein:1999}.
The power spectrum calculated by other codes (e.g., CAMB \cite{Lewis:1999bs})
can also be used as a tabulated input.
The code also calculates the linear growth rate at the output redshift
as a function of wave number $k$, taking into account the effect of
massive neutrinos \cite{Eisenstein:1999}. The linear growth rate is
used to calculate the velocity field later on.

%%%%%%%%%%%%%%%%%%%%%%%%%%%%%%%%%%%%%%%%%%%%%%%%%%%%%%%%%%%%%%%%%%%%%%%%%%%%
\subsubsection{Generating log-normal density field and mock galaxy catalog}
%%%%%%%%%%%%%%%%%%%%%%%%%%%%%%%%%%%%%%%%%%%%%%%%%%%%%%%%%%%%%%%%%%%%%%%%%%%%
Here we briefly summarize the procedure for generating mock galaxy catalogs
using our code (see section \ref{sec:simulation} for further details).
\begin{enumerate}
\item Inverse-Fourier-transform the input power spectrum $P(k)$ to obtain
the two-point correlation function $\xi(r)$; calculate the two-point correlation
function of the Gaussian field $\xi^G(r)$ by \refeq{xiG and xi};
Fourier-transform $\xi^G(r)$ to obtain $P^G(k)$.

\item Generate a Gaussian random field $G(\vk)$ for each cell in Fourier space
from $P^G(k)$, using \refeq{Gk}; Fourier transform $G(\vk)$ to obtain $G(\vx)$
and log-transform $G(\vx)$ to obtain the log-normal density fluctuation $\delta(\vx)$.
The code generates $\delta(\vx)$ for both biased (i.e., galaxy) and unbiased
(i.e., matter) density fields.

\item Generate a velocity field from the matter density fluctuation
using the linear continuity equation (\ref{eq:cont_eq_x}).

\item Generate discrete galaxy positions from the galaxy density fluctuation
by Poisson sampling the density field in each cell with galaxies being
      randomly distributed within the cell.

\end{enumerate}
The code outputs the mock galaxy catalog to a binary file. This file
contains the 3-D positions and velocity components of each galaxy in units of
$\hMpc$ and km/s respectively.

%%%%%%%%%%%%%%%%%%%%%%%%%%%%%%%%%%%%%%%%%%%%%%%%%%%%%%%%%%%%%%%%%%%%%%%%%%%%
\subsubsection{Estimating the power spectrum multipoles}
%%%%%%%%%%%%%%%%%%%%%%%%%%%%%%%%%%%%%%%%%%%%%%%%%%%%%%%%%%%%%%%%%%%%%%%%%%%%
The next step is to estimate the power spectrum from the simulated galaxy catalog.
In our code we use FFT to do this \cite{Feldman:1993}.
FFT requires a local number density of galaxies at regular grid points.
The code has two options for the density assignment scheme;
the Nearest-Grid-Point (NGP) assignment and the Cloud-In-Cell (CIC) assignment.
See \cite{jing:2004} for the details of the effect of density assignment
on the power spectrum estimation with FFT.

What we want to measure is the galaxy power spectrum multipoles $P_{l}(k)$ defined as 
\begin{equation}
P(k,\mu) = \sum_{l=0}^{\infty} P_{l}(k) \mathcal{L}_l(\mu) \,,
\label{eq:defpl}
\end{equation}
where $\mathcal{L}_l$ is the $l$-th order Legendre polynomial.
The two-dimensional power spectrum $P(k,\mu)$ is estimated from the simulated galaxy catalog as 
\begin{equation}
P(k,\mu) = 
W^{-2}_{\rm mesh}(\bm{k})\left[ \dfrac{1}{N_k} \sum^{N_k}_{i=1} |\delta_g(\bm{k}_i)|^2-P_{\text{shot}}\right] \,,
\end{equation}
where $\delta_g(\bm{k})$ is the Fourier transformed local galaxy number
density contrast, $N_k$ is the number of Fourier modes within the given $k$ bin, $P_{\text{shot}}$ is the power spectrum for a uniform random distribution of particles (shot noise), and $\mu$ is the cosine of the angle between $\vk$ and the line-of-sight vector.
Here we assume that the line-of-sight vector is the same for all galaxies (i.e., global plane-parallel approximation).
The mesh window function $W_{\rm mesh}(k)$ is the Fourier transform of the density assignment scheme, that is $W_{\rm mesh}(k) = \left[\sin(kH/2)/(kH/2)\right]^p$,  where $H$ is the mesh size, and $p=1$ and $2$ for NGP and CIC, respectively. The shot noise contribution for NGP and CIC is calculated analytically \cite{djeong_thesis:2010}.
For NGP, if we use the same mesh size as the log-normal galaxy
realisation, we do not need to correct for the density assignment.
This is because the log-normal realisation code also uses FFT and density fields are generated at regular grid points.

To estimate the {\it true} power spectrum multipoles $P_{l}(k)$,
we first compute a brute-force estimate $\hat{P}_m(k)$ as 
\begin{equation}
\hat{P}_m(k) = \sum_{\mu} P(k,\mu) \mathcal{L}_m(\mu) \,.
\end{equation}
Using \refeq{defpl}, $\hat{P}_m(k)$ can be rewritten as
\begin{equation}
\hat{P}_m(k) = \sum_{\mu} \sum_{l=0}^{\infty} P_{l}(k) \mathcal{L}_{l}(\mu) \mathcal{L}_m(\mu)
= \sum_{l=0}^{\infty} P_{l}(k) \mathcal{M}_{lm},
\end{equation}
where $\mathcal{M}_{lm}$ is the $\mu$-leakage matrix defined as 
\begin{equation}
\mathcal{M}_{lm} = \sum_{\mu} \mathcal{L}_{l}(\mu) \mathcal{L}_m(\mu).
\end{equation}
Then the true power spectrum multipoles $P_{l}(k)$ can be calculated as 
\begin{equation}
P_{l}(k) = \sum_{m=0}^{\infty} \hat{P}_m(k) \mathcal{M}^{-1}_{lm},
\end{equation}
where $\mathcal{M}^{-1}_{lm}$ is the inverse matrix of $\mathcal{M}_{lm}$.
Note that $\mathcal{M}_{lm}$ is computed for each $k$ separately.
While the sum over $m$ goes to $\infty$, we only sum up to a certain
maximum value, specified as an input in the configuration file.

Depending on the geometry of the volume, the matrix inversion
of $\mathcal{M}_{lm}$ sometimes becomes unstable particularly at
low $k$ where the number of $\mu$ modes for a fixed $k$ is limited.
To avoid this problem we include another option for the power spectrum
estimation, that is, the ``cubic box mode''.
In this mode the mocked volume (it is rectangular cuboid in general)
is embedded in a cube which is large enough to contain the whole mocked box,
setting $\delta_g(\vx) = 0$ outside of the mock catalog box.
This procedure enables us to have a sufficient number of $\mu$ modes
even at a small $k$.
However, in the cubic box mode, $P_{l}(k)$ is convolved with the survey
window function $W(\bm{r})$: namely, $W(\bm{r}) = 1$ if $\bm{r}$
is within the mock catalog box and $W(\bm{r}) = 0$ otherwise.
We need to correct this survey window effect when comparing with
the theoretical predictions of the power spectrum multipoles.
For the treatment of the survey window function in redshift space, 
we refer the reader to, e.g. \cite{Beutler:2013yhm}.

%%%%%%%%%%%%%%%%%%%%%%%%%%%%%%%%%%%%%%%%%%%%%%%%%%%%%%%%%%%%%%%%%%%%%%%%%%%%
\subsubsection{Test the code}
%%%%%%%%%%%%%%%%%%%%%%%%%%%%%%%%%%%%%%%%%%%%%%%%%%%%%%%%%%%%%%%%%%%%%%%%%%%%
To check the results, run some number of realisations, compute
real-space monopole power spectra, average them, and compare the average with the
input matter power spectrum times the bias squared.
The average should precisely reproduce the input power spectrum on large scales,
although it would deviate from the input on small scales due to the resolution
effect of Fourier meshes. See \reffig{pk_real}.

\bibliography{references}

\end{document}